\documentclass[sigconf]{acmart}
\usepackage{graphicx}
\usepackage{longtable}
\usepackage{array}
\usepackage{booktabs}
\usepackage[table,xcdraw]{xcolor}
\usepackage{float}
\usepackage{ulem}
\usepackage{url}
\usepackage{multirow}
\usepackage{makecell}
\usepackage{tabularx}
\usepackage{soul}
\usepackage{eurosym}
\usepackage{enumitem}
\usepackage{wasysym}
\usepackage{subcaption} 
\usepackage{xspace}
\usepackage{tabularray}
\usepackage{placeins}
\setlist[itemize]{leftmargin=*,labelsep=5pt}
\usepackage{totpages}

\newcommand{\vheader}[1]{\rotatebox[origin=c]{90}{\makecell{#1}}}
\newcommand{\Full}{\CIRCLE}
\newcommand{\Half}{\LEFTcircle}
\newcommand{\Empty}{\Circle}

\newcolumntype{C}[1]{>{\centering\arraybackslash}p{#1}}
\newcolumntype{L}[1]{>{\raggedright\arraybackslash}p{#1}}

\newcommand{\DD}{UMBRA\xspace}

\AtBeginDocument{%
  }

\setcopyright{acmlicensed}
\copyrightyear{2018}
\acmYear{2018}
\acmDOI{XXXXXXX.XXXXXXX}
\acmConference[Conference acronym 'XX]{Make sure to enter the correct
  conference title from your rights confirmation email}{June 03--05,
  2018}{Woodstock, NY}
\acmISBN{978-1-4503-XXXX-X/2018/06}

\begin{document}

\title{When the Abyss Looks Back: Unveiling Evolving Dark Patterns in Cookie Consent Banners}

\author{Nivedita Singh}
\affiliation{%
   \institution{Sungkyunkwan University}
   \city{Suwon}
  \country{South Korea}
}

\author{Seyoung Jin}
\affiliation{%
   \institution{Sungkyunkwan University}
    \city{Suwon}
\country{South Korea}
}
\author{Hyoungshick Kim}
\authornotemark[1]
\affiliation{%
 \institution{Sungkyunkwan University}
  \city{Suwon}
 \country{South Korea}
}



\begin{abstract}
To comply with data protection regulations such as the EU's General Data Protection Regulation (GDPR) and California Consumer Privacy Act (CCPA), websites widely deploy cookie consent banners to collect users' privacy preferences. In practice, however, these interfaces frequently embed dark patterns that undermine informed and freely given consent. As regulatory scrutiny increases, dark patterns have not disappeared; but instead evolved into more subtle and legally ambiguous forms, making existing detection approaches increasingly outdated. We present \DD, a consent management platform (CMP)-agnostic system that detects both previously studied patterns (DP1--DP10) and nine newly evolved patterns (DP11-DP19) targeting information disclosure, consent revocation, and legal ambiguity, including pay-to-opt-out schemes, revocation barriers, and fake opt-outs. \DD combines text analysis, visual heuristics, interaction tracing, and cookie-state monitoring to capture multi-step consent flows that prior tools miss. We evaluate \DD on a manually annotated ground-truth dataset and achieve 99\% detection accuracy. We further conduct a large-scale compliance-oriented measurement across 14,000 websites spanning the EU, the US, and top-ranked global domains. The results show that evolved dark patterns are pervasive: revocation is frequently obstructed, cookies are often set before consent or despite explicit rejection, and opt-out interfaces often fail to prevent third-party tracking. Beyond usability, we show concrete security implications: on sites exhibiting revocation barriers, cookies increase by 25\% on average, and many cookies use insecure attributes that increase exposure to attacks such as XSS and CSRF. Overall, our findings provide evidence of systematic non-compliance and demonstrate how evolving consent manipulation erodes user autonomy while amplifying privacy and security risks. 
\end{abstract}

\maketitle
\section{Introduction}

Whenever users visit a website, a cookie consent banner or dialog typically appears, prompting them to set their preferences regarding third-party cookies. Obtaining consent for accepting or rejecting third-party cookies is valuable for websites, particularly for third-party advertising partners, and is required by GDPR~\cite{eijk2019impact}. 
The GDPR explicitly mandates that consent be freely given, informed, and unambiguous before personal data is processed, thereby placing cookie consent interfaces at the center of practical privacy enforcement~\cite{leiser2023dark}. As a result, cookie consent banners have become a primary mechanism through which regulatory intent is translated into real-world system behavior.

However, many websites exploit this regulatory requirement by employing manipulative or coercive design practices, commonly referred to as dark patterns, that subtly nudge users into agreeing to data collection~\cite{cranor2022cookie}. These tactics often make it difficult to opt out, for example by visually prioritizing the ``accept'' option, obscuring rejection controls, or presenting consent information in overly complex language. Such banners embody adversarial user interface strategies that systematically subvert user intent. Despite more than seven years of GDPR enforcement, these practices remain widespread. Regulators have imposed substantial fines for non-compliant consent mechanisms, including penalties of \$170 million against Google and \$68 million against Facebook~\cite{Google_Fine}, yet cookie consent designs continue to evolve in ways that evade effective compliance.

Recent regulatory guidance further requires websites to provide accessible and symmetric mechanisms for withdrawing consent after it has been given. In practice, however, similar dark patterns frequently persist in consent withdrawal interfaces~\cite{kancherla2025johnny}. These designs undermine user autonomy~\cite{gray2018dark} and potentially violate the GDPR requirement, Article. 7 (3) that withdrawing consent must be as easy as giving it~\cite{Article_7_3,santos2019cookie}. Importantly, these revocation-related patterns do not merely affect user experience; they determine whether tracking continues despite an explicit attempt to opt out, directly shaping post-consent data collection behavior.

\begin{figure}[t]
    \centering
    \includegraphics[width=\linewidth]{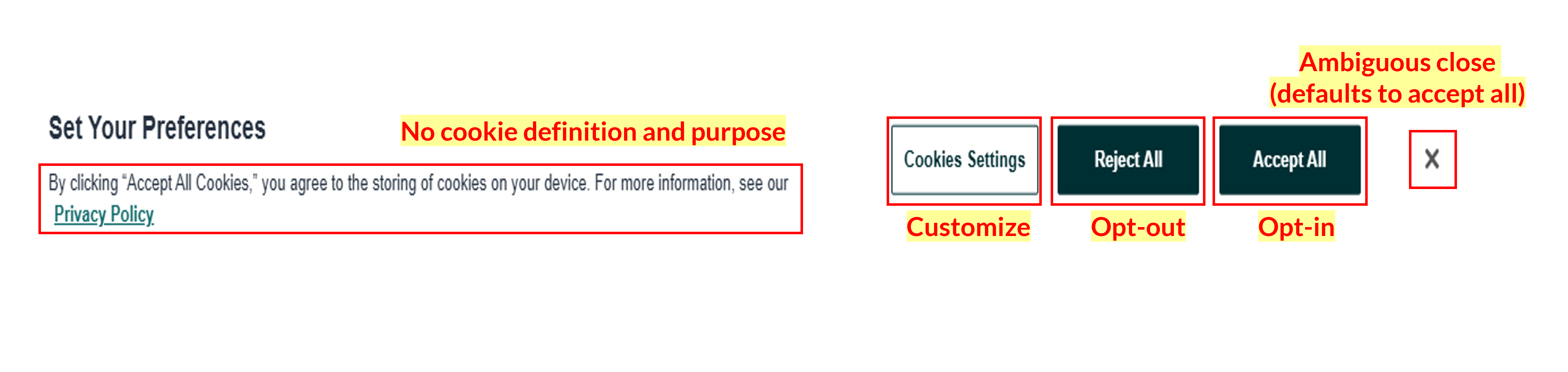}
    \caption{Examples of cookie consent banners featuring dark patterns, such as CookieInfoDisplay (DP11) and PurposeInfoDisplay (DP12) (Source: \url{www.globenewswire.com}).}
    \Description{Examples of cookie consent banners featuring dark patterns, such as CookieInfoDisplay (DP11) and PurposeInfoDisplay (DP12) (Source: \url{www.globenewswire.com}).}
    \label{fig:Bannerexample1}
\end{figure}

\begin{figure}[t]
    \centering
    \includegraphics[width=\linewidth]{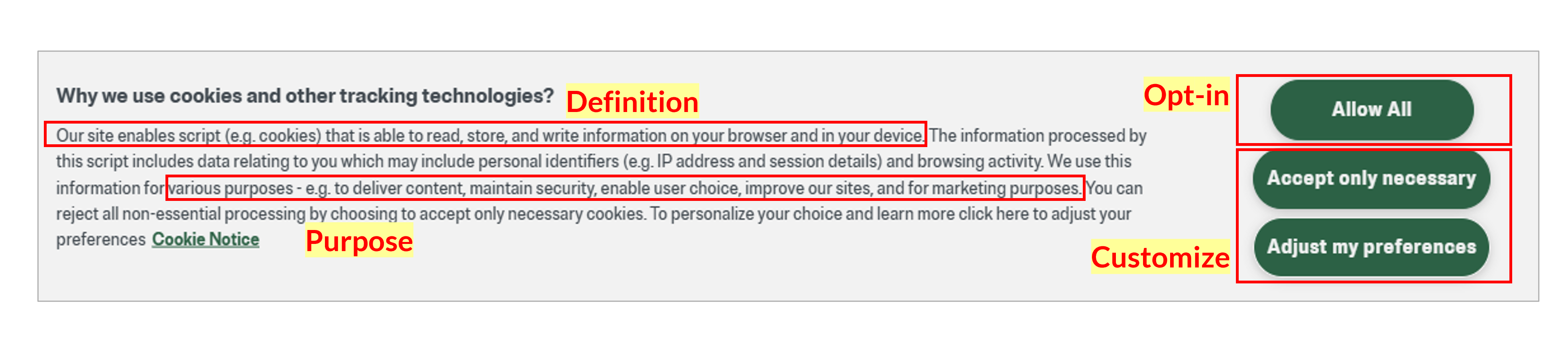}
    \caption{Example of an ideal cookie banner from our research perspective (Source: \url{www.onetrust.com}).}
     \Description{Example of an ideal cookie banner from our research perspective (Source: \url{www.onetrust.com}).}
    \label{fig:Bannerexample2}
\end{figure}

Previous studies have taken important steps toward measuring dark patterns in cookie consent banners, primarily focusing on a fixed set of well-known patterns (DP1--DP10)~\cite{kirkman2023darkdialogs, bouhoula2024automated}. However, these approaches are increasingly outdated due to the rapid evolution of consent interfaces. Modern banners adopt more subtle and legally ambiguous strategies, such as multi-step consent revocation flows and pricing-based coercion mechanisms like ``OptOutPricing (DP13)''  and ``Revocation Hard (DP15),'' which are not captured by existing detectors. As a result, prior tools fail to identify a growing class of manipulative designs that emerge only through interaction and layered interfaces.

Moreover, existing work largely treats dark patterns as usability or regulatory compliance issues and does not systematically examine their security and privacy consequences. In particular, prior studies do not analyze how manipulative consent designs affect cookie-setting behavior in practice. This omission is critical, as dark patterns can enable the deployment of excessive third-party cookies and trackers, thereby increasing exposure to security risks arising from insecure cookie policies~\cite{kancherla2025johnny}. Without linking interface-level manipulation to concrete system behavior, the real-world impact of evolved dark patterns remains poorly understood.

To address these gaps, we present \DD, a Consent Management Platform (CMP)-agnostic system detecting both DP1–DP10 and nine newly evolved patterns (DP11--DP19), including revocation barriers and fake opt-outs. \DD combines text analysis, visual heuristics, interaction-based monitoring, and cookie-state analysis to capture complex consent flows that manifest only after user interaction. Using \DD, we conduct the first compliance-oriented, large-scale measurement study of evolved dark patterns, guided by the following research questions:

\textbf{RQ1.} Can a CMP-agnostic detector accurately identify both established (DP1--DP10) and newly evolved dark patterns (DP11--DP19) across diverse cookie consent banners?

\textbf{RQ2.} How widespread are newly evolved dark patterns (DP11--DP19) across different regions and website categories?

We summarize our contributions as follows:
\begin{itemize}
    \item \textbf{Enhanced detection of evolved dark patterns.} We extend existing detectors by incorporating nine newly evolved dark patterns (DP11--DP19) that capture contemporary consent manipulation strategies, including revocation barriers and pricing-based coercion.
    \item \textbf{Security-aware cookie-state analysis.} Beyond interface detection, \DD monitors cookies across different interaction states. We show that websites exhibiting revocation Hard (DP15) set 25\% more cookies on average, many of which increase security and privacy risks.
    \item \textbf{CMP-agnostic detection framework.} We design a CMP-agnostic framework that integrates dialog extraction, clickable classification, text and visual analysis, and cookie-state monitoring.
    \item \textbf{High detection accuracy at scale.} We evaluate \DD on a manually annotated dataset, achieving 99\% accuracy, and conduct a large-scale measurement across 14,000 websites from the EU, the US, and top-ranked global domains. Prior studies reported accuracy of 91\%~\cite{eijk2019impact}.
    \item \textbf{Evidence of widespread non-compliance.} Our study provides the first comprehensive compliance-oriented evidence that evolved dark patterns are pervasive and closely linked to increased tracking and insecure cookie practices.
\end{itemize}

Overall, our findings demonstrate that consent manipulation remains systematic and increasingly sophisticated, eroding user autonomy while amplifying tangible security and privacy risks. By bridging interface-level detection with security-aware measurement, this work provides actionable insights for researchers, regulators, and practitioners seeking to enforce meaningful data protection on the modern web.
\section{Background \& Threat Model}

This section establishes the foundational context for our study. We first review the evolution of privacy regulations and third-party cookies, followed by an examination of manipulative design practices known as dark patterns. Finally, we formalize the threat model, characterizing the adversarial capabilities of non-compliant websites and the resulting security risks to users.

\subsection{Privacy Regulation and Third-Party Cookies}
Originally designed to maintain stateful sessions for first-party websites, HTTP cookies have evolved into a pervasive infrastructure for cross-site tracking and targeted advertising~\cite{lin2024browsing,singh2025crumbled}. This transformation turned user data into a commodified asset, often harvested without the user's knowledge~\cite{hu2019characterising}. To curb this unregulated surveillance, legislative frameworks such as the EU's ePrivacy Directive and the GDPR, as well as the US CCPA, imposed strict consent requirements~\cite{bond2012eu, GDPR1, CCPA1}. Under these regimes, third-party cookies—particularly those used for profiling—are treated as personal data identifiers, legally mandating explicit, prior consent from the user before any data collection occurs.

\subsection{Privacy Regulation and User Consent}
The cornerstone of modern privacy law is the principle that consent must be informed, explicit, and freely given~\cite{UserConsnet1}. Regulations mandate that users be empowered to make granular choices about their data without being misled or coerced~\cite{matte2020cookie}. Consequently, cookie consent banners have become the primary regulatory checkpoint on the web. However, a significant gap exists between legal intent and technical implementation. Driven by the economic incentive to maximize data yield, many websites have adopted adversarial interface designs that exploit cognitive biases, systematically nudging users toward ``acceptance'' while obfuscating opt-out mechanisms~\cite{nouwens2020dark, utz2019informed, giese2022factors, borberg2022so, soe2022automated, ahuja2025towards}.

\subsection{Dark Patterns on Consent Banners}
Dark patterns in this context are defined as user interface design choices crafted to subvert user autonomy and manipulate consent decisions. These patterns prioritize the commercial interests of service providers and advertisers over user rights. While early dark patterns relied on simple visual deception—such as highlighting ``Accept All'' buttons or hiding ``Reject'' options~\cite{bielova2024effect, kirkman2023darkdialogs}—recent trends show an evolution toward more sophisticated and legally ambiguous tactics. These include creating friction in the revocation process, employing pay-to-opt-out schemes, or using confusing legal terminology to manufacture consent~\cite{morel2025will}. Such designs not only violate the spirit of ``freely given'' consent but also actively degrade the trustworthiness of the web ecosystem.

\subsection{Threat Model}
We define the threat model for cookie consent manipulation as follows:

\noindent\textbf{Adversary.} The adversary is the website operator or the CMP. Their objective is to maximize the consent rate (opt-in) to sustain ad-revenue models while minimizing the apparent cost of compliance. They possess full control over the website's DOM, the banner's logic, and the cookie-setting mechanisms.

\noindent\textbf{Attack vector.} The primary vector is the consent interface (banner). The adversary employs \textit{interface interference} (dark patterns) to exploit the user's limited attention and ``System 1'' thinking (fast, automatic), thereby bypassing regulatory protections like the GDPR and CCPA.

\noindent\textbf{Victim.} The victim is the website visitor. They are often unaware of the underlying tracking mechanisms and assume that the interface accurately reflects their choices.

\noindent\textbf{Consequences.} The successful execution of these dark patterns results in:
\begin{itemize}
    \item \textbf{Privacy Violation:} Unauthorized collection of personal data and cross-site profiling despite the user's intent to opt-out.
    \item \textbf{Security Exposure:} Coerced consent often authorizes the loading of unvetted third-party scripts. As demonstrated in our findings, this significantly increases the attack surface, exposing users to security vulnerabilities such as Cross-Site Scripting (XSS) and Cross-Site Request Forgery (CSRF) from compromised advertising networks.
    \item \textbf{Regulatory Non-Compliance:} The system fails to meet the legal standards of revocability and informed consent, creating legal liability.
\end{itemize}
\section{\DD Design}
We present \DD, a CMP-agnostic system designed to uncover regulatory evasion tactics that have evolved beyond the scope of prior detection tools~\cite{kirkman2023darkdialogs}. Unlike predecessors that focus on static interface properties, \DD incorporates a multi-stage analysis pipeline---spanning text analysis, visual heuristics, DOM interaction tracing, and stateful cookie monitoring---to identify nine novel dark patterns (DP11--DP19). These patterns specifically target information asymmetry, barriers to consent revocation, and deceptive legal framing. The system is structured into five core modules, described below.

\subsection{Dialog Extraction and Ranking}
\DD initiates analysis by identifying cookie consent banners using a hybrid approach that combines CSS selector lists, custom keyword-based heuristics, and visual dominance checks. To handle false positives, the system employs a scoring function that evaluates candidate elements based on text density, z-index visibility, and structural uniqueness. The highest-ranked candidate is classified as the active cookie banner for subsequent analysis.

\subsection{Clickable Classification}
To facilitate interaction, \DD classifies UI elements into functional categories: \textit{Opt-in, Opt-out, More Options, Preference Sliders, Close Buttons,} and \textit{Policy Links}. We employ a dual-method approach: deterministic keyword matching (e.g., ``accept,'' ``manage preferences'') combined with DOM-based structural analysis. To address the linguistic evolution of consent banners---such as the use of informal phrases like ``I am ok''---we expanded the standard dictionary to include colloquialisms observed during our pilot study. The system also supports dynamic lexicon updates to adapt to emerging interface trends.

\subsection{Cookie Setting Behaviour}
A core contribution of \DD is its stateful monitoring of cookie lifecycle events. The system captures cookie storage (both first- and third-party) across four distinct interaction states: \textit{Initial Load, Opt-in, Opt-out,} and \textit{Close}. By performing a differential analysis of these states, \DD detects covert non-compliance, such as Pre-Consent Cookies (DP16) or the persistence of tracking identifiers despite an explicit rejection (Fake Opt-Out, DP18). Persistent third-party cookies are further analyzed to identify ID-like tracking vectors.

\subsection{Dark Pattern Detection Heuristics}
\DD encodes dark patterns into deterministic operational heuristics. We retain the 10 baseline patterns (DP1--DP10) from prior literature~\cite{kirkman2023darkdialogs} for comparability, while introducing nine novel patterns (DP11--DP19) derived from recent regulatory guidelines.

\subsubsection{Baseline patterns (DP1--DP10)}
We re-implemented the following ten patterns using established thresholds:

\begin{itemize}
    \item OnlyOptIn (DP1): Presence of an ``opt-in'' button without a corresponding ``opt-out'' or ``more options'' mechanism.
    \item HighlightedOptIn (DP2): Visual dominance of the opt-in path, quantified via grayscale contrast differences (threshold $\approx$170).
    \item ObstructWindow (DP3): The dialog obscures $>60\%$ of the viewport.
    \item ComplexText (DP4): Text complexity yields a Flesch Reading Ease score $\leq 50$.
    \item MoreOptions (DP5): Users are forced into a secondary layer to find rejection options.
    \item AmbiguousClose (DP6): Presence of a ``close'' button alongside ``accept,'' creating ambiguity regarding the consent state.
    \item MultipleDialogs (DP7): Simultaneous display of multiple distinct consent interfaces.
    \item PreferenceSlider (DP8): Consent toggles are pre-enabled (privacy-intrusive default).
    \item CloseMoreCookies (DP9): Closing the banner triggers cookie placement exceeding the initial state.
    \item OptOutMoreCookies (DP10): Explicit opt-out actions fail to prevent additional cookie placement.
\end{itemize}

\subsection{Detection Heuristics For Newly Evolved Patterns (DP11--DP19)}\label{subsec:newDPs}





We extend the baseline detector with nine patterns grounded in regulatory guidance and current CMP practices. Each pattern is operationalized with deterministic rules over first-page text, visual elements, DOM interactions, and cookie-state comparisons across four stages (Initial, Opt-In, Opt-Out, Close).

\subsubsection{CookieInfoDisplay (DP11).} True if the first page lacks a plain-language definition of cookies. We normalize the dialog text to lowercase and check against a curated lexicon of 12 definitional phrases (e.g., ``cookies are small text files,'' and ``cookies contain''). This lexicon was derived from official CNIL~\cite{CNIL} and ICO~\cite{ICO} guidance to capture descriptive transparency rather than mere mentions. This approach operationalizes GDPR Articles 12 and 13, ensuring consent interfaces use clear and intelligible language (see Appendix~\ref{Appendix:Lexicons} for the full lexicon).

\paragraph{Legal justification for DP11} This dark pattern directly corresponds to the transparency and informed consent obligations established in GDPR Articles 12 (1)~\cite{GDPR_Article_12} and 13 (1) (a–c)~\cite{GDPR_Article_12_1}, reinforced by Recital 39~\cite{Recital_39} and 58~\cite{Recital_58}. Article 12 (1) explicitly requires that information provided to users be \textit{``concise, transparent, intelligible and easily accessible, using clear and plain language.''} Recital 39 further emphasizes that \textit{``natural persons should be made aware of risks, rules, safeguards, and rights''} and that processing should be \textit{``transparent to natural persons.''} Likewise, Recital 58~\cite{Recital_58} underlines that \textit{``information addressed to the public or the data subject should be concise, easily accessible and easy to understand, and clear and plain language should be used.''}
DP11 operationalizes these legal provisions by detecting banners that fail to include a plain-language definition of cookies, thereby identifying non-transparent interfaces that hinder informed consent.
\begin{figure}[ht]
    \centering
    \begin{subfigure}[t]{0.85\linewidth}
        \centering
        \includegraphics[width=\linewidth]{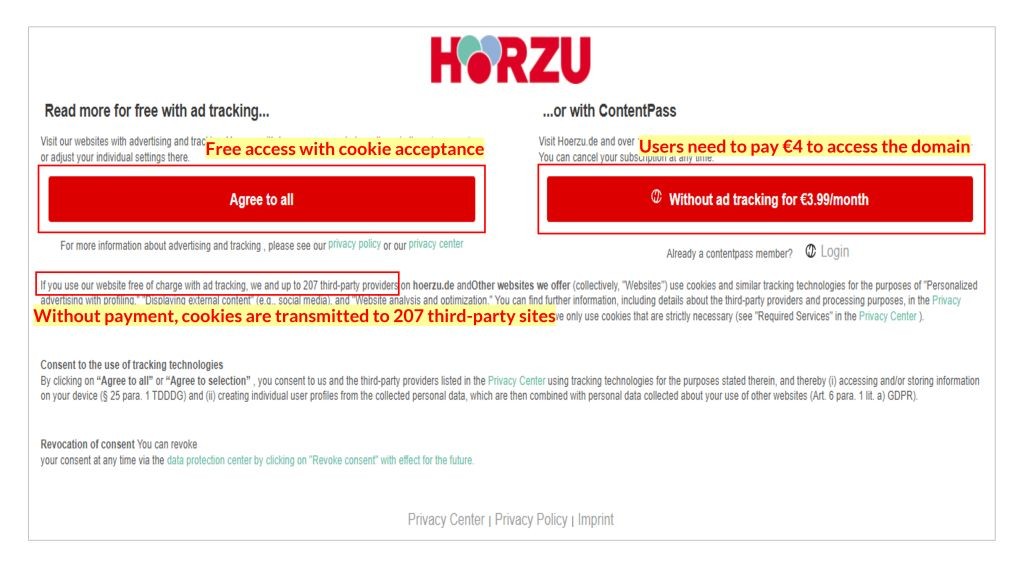}
        \caption{Example of a cookie banner featuring OptOutPricing (DP13): It asks for \euro 4 to opt-out, and if the user declines, profiling may be done by 207 third parties (Source: \url{www.hoerzu.de}).}
        \Description{Example of a cookie banner featuring OptOutPricing (DP13): It asks for \euro 4 to opt-out, and if the user declines, profiling may be done by 207 third parties (Source: \url{www.hoerzu.de}).}
        \label{fig:banner_DP13(1)}
    \end{subfigure}
    \vspace{0.8em} 
    \begin{subfigure}[t]{0.85\linewidth}
        \centering
        \includegraphics[width=\linewidth]{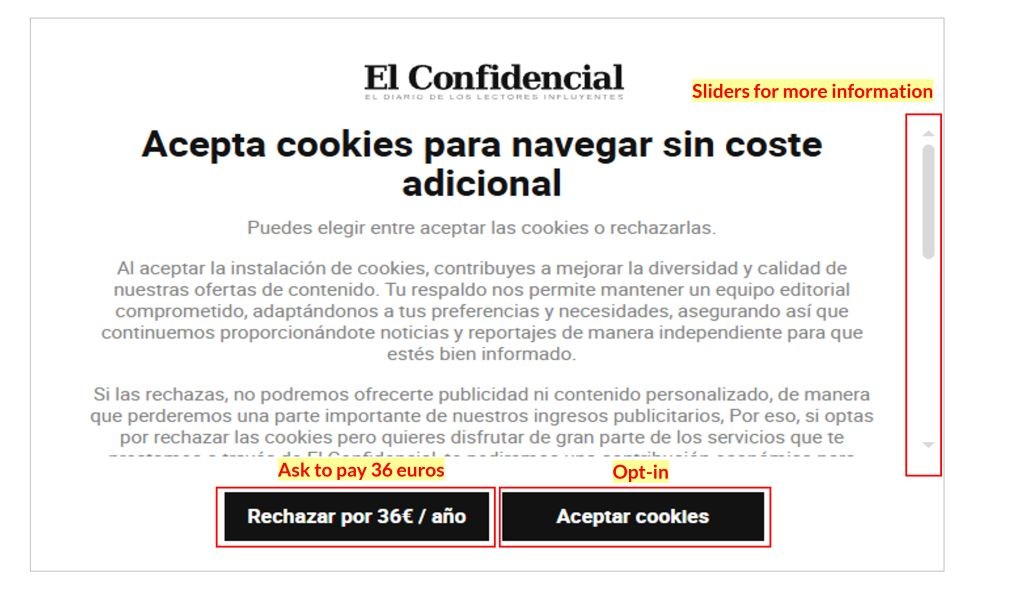}
        \caption{Example of a cookie banner featuring OptOutPricing (DP13): It asks for \euro 36 to opt-out (Source: \url{www.elconfidencial.com}.)}
        \Description{Example of a cookie banner featuring OptOutPricing (DP13): It asks for \euro 36 to opt-out (Source: \url{www.elconfidencial.com}.)}
        \label{fig:banner_DP13(2)}
    \end{subfigure}
    \caption{Example of a cookie banners with DP13.}
    \Description{Example of a cookie banners with DP13.}
     \label{fig:banner_DP13_both}
\end{figure}

\paragraph{Practical Implementation} Although  DP11 is not explicitly mandated in all cases, providing a clear, plain-language definition of cookies in consent banners is crucial, particularly for non-technical users who may not understand what cookies are, such as those seen in widely-used CMPs like OneTrust (see Figure~\ref{fig:Bannerexample2}),  This practice directly aligns with the GDPR’s transparency requirements, Article 12 (1)~\cite{GDPR_Article_12}, ensuring that all users, regardless of their technical knowledge, can easily comprehend what cookies are and how they function. By doing so, it improves the quality of informed consent and contributes to a more transparent, accessible, and user-friendly consent experience.
\subsubsection{PurposeInfoDisplay (DP12).}
This dark pattern identifies banners that omit the purposes of data collection on the initial screen (see Figure~\ref{fig:Bannerexample1}). To detect this, we utilize a lexicon of 50 purpose-expressive phrases curated from CNIL (2020)~\cite{CNIL}, ICO (2023)~\cite{ICO}, and public policy corpora. These phrases are organized into six semantic clusters: analytics, personalization, advertising, security, functionality, and commerce. To handle linguistic diversity, we incorporate fuzzy semantic matching alongside deterministic keyword detection. This enables the recognition of paraphrased or localized expressions (e.g., ``to analyze site traffic'' $\approx$ ``for audience measurement''). This hybrid approach improves robustness, ensuring that the absence of purpose disclosure is not masked by minor rewordings (see Appendix~\ref{Appendix:Lexicons} for full lexicons).

\paragraph{Legal justification for DP12} This dark pattern directly corresponds to the purpose-specification and transparency obligations established in GDPR Article 5 (1) (b)~\cite{GDPR_Article_12_1} and Article 13 (1) (c)~\cite{GDPR_Article_12_1}, reinforced by Recital 39~\cite{Recital_39}. Article 5 (1) (b)~\cite{GDPR_Article_12_1} states that personal data shall be ~\textit{``collected for specified, explicit and legitimate purposes and not further processed in a manner incompatible with those purposes.''} Similarly, Article 13 (1) (c)~\cite{GDPR_Article_12_1} mandates that controllers inform users about \textit{``the purposes of the processing for which the personal data are intended.''} Recital 39~\cite{Recital_39} further emphasizes that individuals must be aware of \textit{``the purposes for which personal data are processed''} to ensure fair and transparent handling.
DP12 operationalizes these provisions by detecting banners that fail to clearly state the purposes of cookies. 

\subsubsection{OptOutPricing (DP13).} This dark pattern, also known as ``pay-to-opt-out,'' occurs when cookie banners require users to pay a fee to reject third-party cookies and access the webpage (see Figure~\ref{fig:banner_DP13_both}). Our tool is designed to detect the co-occurrence of pricing terms (e.g., pay, subscribe) and consent-related keywords (e.g., accept, reject) within the same or closely related textual elements.

\paragraph{Legal justification for DP13}
This dark pattern corresponds to the GDPR prohibition on conditional consent, established under Article 7 (4)~\cite{GDPR_Article_12_1} and Recital 42~\cite{Recital_42}, which require that consent be ``freely given'' and that refusal must not result in a detriment to the user. 
The European Data Protection Board (EDPB) Guidelines~\cite{EDPB_Guidelines} explicitly state that paywalls making access conditional on consent are non-compliant with the GDPR. 
DP13 operationalizes this principle by algorithmically identifying banners in which consent is linked to monetary or contractual conditions (e.g., ``Subscribe or accept cookies,'' ``Accept cookies to continue reading,'' etc.). Such patterns represent potential violations of the principle of freely given consent and undermine user autonomy and transparency in the consent process.

\paragraph{Legal controversy}
While most regulators and scholars argue that ``pay-to-opt-out'' models are in direct conflict with GDPR’s ``freely given consent'' requirement, some have attempted to justify the practice~\cite{morel2025will}. For example
``Premium Content Providers'' argue that offering an alternative ``paid, ad-free experience'' is an acceptable trade-off for users, as it provides a ``clear, transparent choice.''
However, ``regulatory bodies'' have emphasized that consent should not be seen as a ``bargaining tool.'' Consent obtained in this manner would be considered ``coerced,'' and thus ``invalid'' under GDPR, as it cannot be freely given when the user is faced with an ``involuntary financial consequence''~\cite{PayorAccept, PayorAccept1}.
In practical terms, ``cookie paywalls'' create a ``legal grey area'' in the context of the GDPR, as they force users into a ``choice between privacy (rejecting cookies)''  and ``access (paying for content).'' This results in ``unbalanced decision-making'' and undermines the ``informed nature of consent.''
In our study, we encountered numerous cookie banners where the pricing for rejecting cookies or opting out varied significantly across websites (see Figure~\ref{fig:banner_DP13(2)}). This inconsistency highlights the challenge of validating the legality of cookie rejection fees, as no specific regulation governs fixed prices. Without clear legal guidelines on consent fees, the practice remains open to interpretation, allowing websites to potentially exploit this ambiguity and undermine GDPR's principles of free consent.

\subsubsection{ConsentRevocationImpossible (DP14)} 
Websites implement consent revocation in two ways: by offering a method to change preferences at the bottom of the page, or by providing an icon at the far left or right of the webpage. The tool detects the dark pattern as True if no consent revocation is discoverable in the main DOM or iframes via keyword XPath (cookie settings, manage cookies, revoke/withdraw consent) or a badge/icon. The detection of DP14 is achieved through the following key steps: 
\begin{itemize}
    \item Keyword XPath search: The system searches the DOM and iframes for the presence of specific keywords for revocation. If none of the elements are found, DP14 is flagged.
    \item Badge/Icon heuristic: A visual heuristic is employed to detect the presence of icons or badges in the bottom corners of the page, which are commonly used for consent revocation controls. These elements are often circular, a cookie icon, or a CMP logo icon. If such icons are missing, too, DP14 is triggered.
   If any of these terms is found in the DOM or in an iframe, DP14 is not triggered. 
\end{itemize}
\begin{figure}[t]
    \centering
    \includegraphics[width=\linewidth]{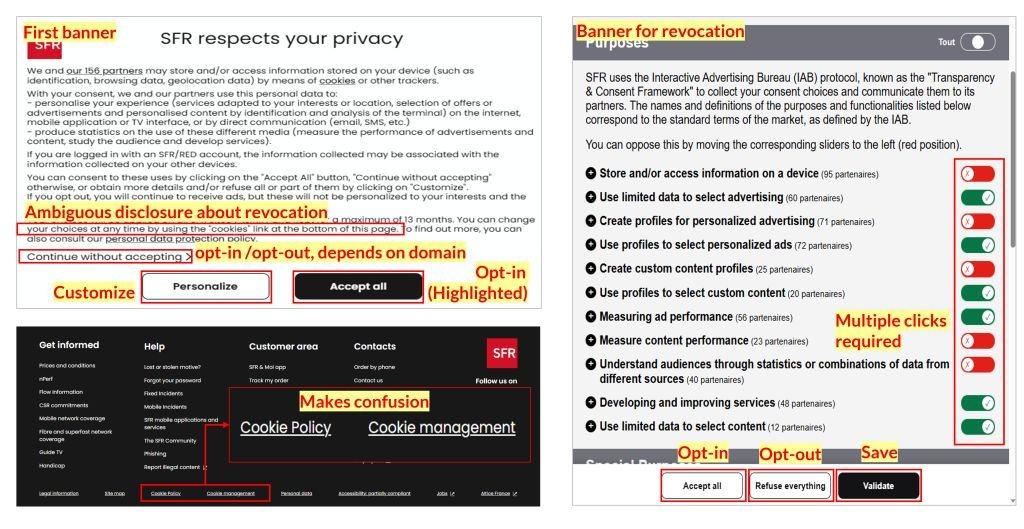} 
    \caption{Example of the cookie banner featuring revocation-Hard (DP15): While revocation is available (at the footer of the webpage), it is not as easy as obtaining consent (Source: \url{www.sfr.fr)}.}
    \Description{Example of the cookie banner featuring revocation-Hard (DP15): While revocation is available (at the footer of the webpage), it is not as easy as obtaining consent (Source: \url{www.sfr.fr)}.}
    \label{fig:DP15example1}
\end{figure}

\begin{figure}[t]
    \centering
    \includegraphics[width=\linewidth]{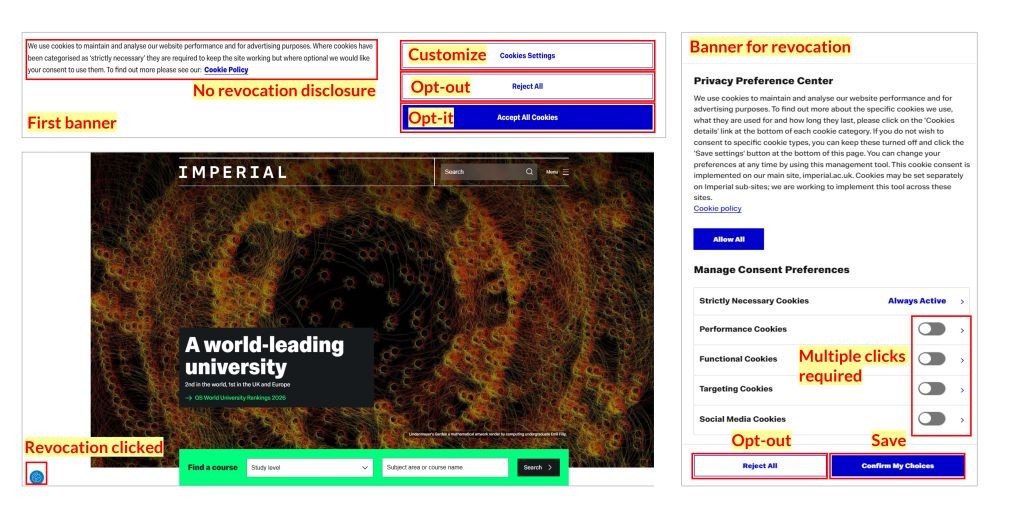} 
    \caption{Example of the cookie banner featuring revocation-Hard (DP15): While revocation is available (cookie icon on the far left of the webpage), the revocation disclosure is not provided on the first banner (Source: \url{www.imperial.ac.uk}).}
    \Description{Example of the cookie banner featuring revocation-Hard (DP15): While revocation is available (cookie icon on the far left of the webpage), the revocation disclosure is not provided on the first banner (Source: \url{www.imperial.ac.uk}).}
    \label{fig:DP15example2}
\end{figure}


\paragraph{Legal justification for DP14}
This dark pattern violates the requirement that consent must be withdrawable at any time and ``as easy to withdraw as to give,'' as established in GDPR Article 7 (3) and reinforced by Recital 42~\cite{Recital_42}.
The EDPB guidelines~\cite{EDPB_Guidelines, Article_7_3}
emphasize that users must be able to revoke consent without detriment and without navigating complex menus. DP14 detects banners without an accessible withdrawal mechanism, whether visible in the DOM, hidden in iframes, or represented by an icon, identifying non-compliant cookie interfaces.

\subsubsection{RevocationHard (DP15).} This pattern is triggered when withdrawing consent becomes difficult, despite the existence of a revocation mechanism, thereby violating Article 7 (3)~\cite{Article_7_3} of the GDPR, which mandates that \textit{revocation be as easy as giving consent}. 

We draw on the four conditions outlined by the EDPB guidelines~\cite{UserConsnet1} to determine whether revocation is considered ``hard'' or not.
If any of the following conditions hold, revocation is classified as hard:
\begin{itemize}
    \item Missing disclosure (C1): First page lacks an explicit ``you may withdraw/change consent at any time'' statement. 
    \item Entry steps (C2): The easiest revocation entry requires $>2$ steps (e.g., hidden in footer or inside an iframe). 
    \item Active click-through (C3): Completing withdrawal (e.g., toggles + save/confirm) requires $>2$ interactions.
    \item Asymmetry (C4): Withdrawing requires strictly more interactions than granting consent on the same site.
\end{itemize}

DP15 is flagged if any of the conditions C1–C4 are met, indicating that consent withdrawal is unnecessarily complicated (see Figure~\ref{fig:DP15example1} and Figure~\ref{fig:DP15example2}).

\paragraph{Legal justification for DP15}
This dark pattern directly corresponds to the GDPR's principle of freely given and informed consent as defined under Article 7 (3) and Recital 42~\cite{Recital_42}. Article 7 (3)~\cite{EDPB_Guidelines, Article_7_3, Article7_2} clearly states that \textit{``The data subject shall have the right to withdraw his or her consent at any time''} and further mandates that \textit{``It shall be as easy to withdraw as to give consent.''} Recital 42~\cite{Recital_42} reinforces this by specifying that \textit{``Consent should not be regarded as freely given if the data subject is unable to refuse or withdraw consent without detriment.''}
Accordingly, DP15 operationalizes this legal requirement by detecting banners that impose unnecessary friction or obstacles in the withdrawal process—such as multi-step revocation flows, buried settings.

\subsubsection{PreConsentCookies (DP16).} 
This dark pattern is triggered if cookies are set before any user action at \textit{Initial}. For detection, we collected and identified the cookies when the webpage loaded, before the user made any selections on the cookie banner. In \textit{standard} mode we require any third-party cookies; in \textit{strict} mode we also count ID-like cookies. 
DP16 is triggered when cookies are set before the user has provided consent. The detection involves the following steps:
\begin{itemize}
    \item Cookie collection: Collects data on cookies set during the initial page load.
    \item Pre-consent check: If cookies are set before consent is obtained, DP16 is flagged.
    \item Strict mode: In strict mode, all cookies, including first-party, are considered.
\end{itemize}
ID-like cookies are those that can be used for tracking, specifically those that help build user profiles for targeted advertising. They are often persistent and can be linked to an identifiable user across different websites.

\paragraph{Legal justification for DP16}
This dark pattern directly corresponds to the GDPR principle of lawfulness and fairness in data processing, as established in Article 5 (1) (a)~\cite{Article_5}, which states that personal data should be processed ``lawfully, fairly, and in a transparent manner in relation to the data subject.'' According to Recital 30~\cite{Recital_30}, data processing ``\textit{should not be regarded as lawful}'' if consent has not been obtained prior to processing. Specifically, cookies should not be set before a user consents, as they fall under the category of personal data processing~\cite{Article_4, Recital_30}.

Furthermore, Article 7 (1)~\cite{Article_7_3} \cite{Article7_2} of the GDPR mandates that consent must be ``\textit{freely given, specific, informed, and unambiguous.}'' The absence of consent before cookies are set undermines the GDPR's transparency requirement. Recital 30~\cite{Recital_30} clarifies that data subjects should not face any detriment (e.g., limited functionality or forced tracking) if they choose to refuse consent. Therefore, DP16 operationalizes this requirement by detecting cookies set prior to user consent.


\subsubsection{LegalAmbiguity (DP17).} 
Triggered when the banner fails to disclose a clear, GDPR-accepted legal basis for data processing. If the banner has a second layer (e.g., ``Privacy Policy''), the check is re-applied after it is opened to ensure the legal basis is properly disclosed. 
To detect this dark pattern, we used the following two terms.
\begin{itemize}
\item Activity terms: The system analyzes the text of the first banner to identify if user data processing is mentioned, such as ``process personal data,'' ``user tracking,'' etc. If such terms are found, the system then checks for second-layer links such as ``Privacy Policy,'' ``Data Protection Regulation,'' or ``More Options'' to verify the presence of legal basis terms.
\item Legal basis terms: The system navigates to subpages to search for legal basis terms for data processing, such as ``legitimate interest (GDPR),''  ``Article 6,''  ``California Notice at Collection'' etc.
\end{itemize}
If the activity term is present but the legal term is missing, DP17 is triggered as true; otherwise, it is not.

\paragraph{Legal justification for DP17}
GDPR Articles 6 and 13~\cite{Article_6, GDPR_Article_13} require that personal data processing be based on a clear legal basis, which must be disclosed to users. Article 6 outlines several lawful bases for processing, including consent, legitimate interest, contractual necessity, legal obligation, and public task. Article 13 mandates that this legal basis be communicated to users at the time of data collection. Recital 39~\cite{Recital_39} also addresses the concept of legal ambiguity in relation to data processing.


This ambiguity in the legal basis violates GDPR’s requirement for transparency, clarity, and explicit communication of processing purposes and lawful grounds.

\subsubsection{FakeOptOut (DP18).} 
This pattern is triggered if cookies persist even after selecting opt-out.
DP18 is triggered when the ``opt-out'' option fails to stop cookies from being set after a user chooses Opt-Out. If cookies are set despite rejection, it indicates that the opt-out mechanism is non-functional. The detection involves the following steps:
\begin{itemize}
    \item Post-rejection cookies: The system checks whether any cookies are still set after the user selects ``Opt-Out.'' If DP18 is triggered, we collect the cookies for further analysis.  

\end{itemize}

\paragraph{Legal justification for DP18}
GDPR Article 7  (4)~\cite{Article_7_3} and Recital 43~\cite{Recital_43} clearly establish that consent must be freely given and withdrawn without detriment. A fake opt-out pattern, in which the rejection mechanism is non-functional (i.e., cookies remain set after the user selects ``Reject''), directly contradicts this principle. Users must be able to reject cookies without being penalized with unwanted data processing. Article 7 (4)~\cite{Article_7_3, Article7_2} states, \textit{``When assessing whether consent is freely given, the consequences of refusing consent should not be disproportionate,''} while Recital 43~\cite{Recital_43} emphasizes, \textit{``A refusal to consent should not result in a detriment.''} Additionally, Article 5 (1) (a)~\cite{Article_5} mandates that \textit{``Personal data shall be processed lawfully, fairly, and in a transparent manner in relation to the data subject.''} The fake opt-out pattern violates the principle of consent being freely given and not conditional upon acceptance of cookies.
DP18 operationalizes the GDPR requirements for freely given consent by identifying cookie banners where the rejection option fails to prevent cookies from being set.
\subsubsection{MultiClickOptOut (DP19).}
This dark pattern is triggered if the measured clicks to opt out exceed the clicks to opt in, or if no opt-out/more-options is provided while an opt-in is present. We log per-site click asymmetry as evidence. DP19 is triggered when the opt-out process requires more clicks or is less visible than the opt-in process. The detection involves the following steps:
\begin{itemize}
    \item Opt-in clicks: Counts the number of clicks needed to opt-in (typically ``Accept'' or ``Allow'').
    \item Opt-out clicks: Counts the number of clicks needed to opt-out, which may involve additional steps.
    \item Comparison: If the opt-out process requires more clicks or is less visible than the opt-in process, DP19 is flagged.
\end{itemize}

\paragraph{Legal justification for DP19}
This dark pattern addresses the asymmetry in the number of clicks required to opt in versus opt out. If the opt-out process requires more clicks than the opt-in process or if the opt-out option is not clearly visible, DP19 is triggered. This violates GDPR Article 7 (4)~\cite{Article_7_3, Article7_2} and Recital 42~\cite{Recital_42}.
GDPR Article 5 (1) (a)~\cite{Article_5} further requires that, if the opt-out mechanism is more complex or less visible than the opt-in process, users are coerced into accepting cookies, undermining the principle of freely given consent. DP19 operationalizes this requirement by flagging when the opt-out process is unnecessarily complicated or hidden, which violates the GDPR's core consent principles.
\subsection{Rule Formation And Detection Criteria} Since \DD operates on heuristic- and rule-based logic, every detection criterion was precisely defined through iterative discussions between the lead and second researcher. Each rule was established only after both researchers reached full agreement on its operational validity and threshold conditions.

For patterns such as DP11, DP12, DP13, and DP17 the keyword lexicons were finalized after a thorough analysis of hundreds of cookie banners and cross-referencing with regulatory documents from the CNIL, ICO, and EDPB guidelines. 
During this process, both researchers critically evaluated candidate keywords to avoid overly generic or context-independent terms that could yield false positives. Only context-specific expressions with direct relevance to cookie transparency or lawful processing were retained in the final rule set.

For example, in defining DP15, the researchers agreed to ground the detection logic strictly in EDPB Guidelines~\cite{UserConsnet1} on Consent, rather than in subjective or layman interpretations of  ``difficulty.'' Accordingly, consent revocation was classified as hard only when the banner demonstrably violated one or more of the guideline-based conditions. 

Each detection rule thus encapsulates a balance between regulatory compliance, linguistic precision, and technical reproducibility, ensuring that \DD’s automated decisions remain both interpretable and legally traceable.


\section{Methodology and Work Flow} 
\begin{figure}[t]
    \centering
    \includegraphics[width= 1\linewidth]{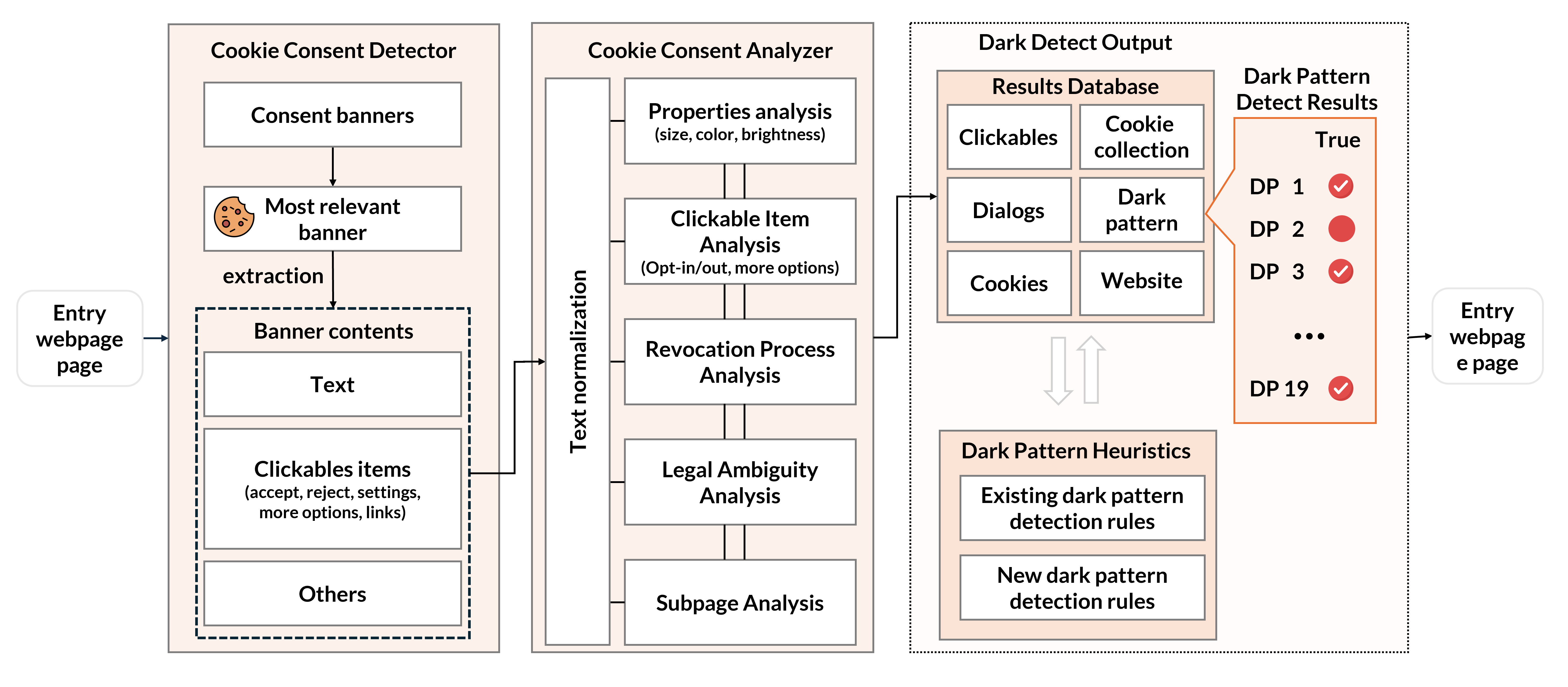} 
    \caption{Overview of our work: The figure illustrates the process, from loading the webpage to detecting dark patterns and saving the results.}
    \Description{Overview of our work: The figure illustrates the process, from loading the webpage to detecting dark patterns and saving the results.}
    \label{fig:Overview}
\end{figure}

Figure~\ref{fig:Overview} illustrates the workflow of our approach. The tool begins by receiving the URL and detecting the cookie banner. In the first stage, it selects the most relevant banner and extracts its elements, such as the banner text and clickable items. In the second stage, the system analyzes the cookie consent banner for dark pattern detection. The text is normalized by detecting non-english content and converting it into English, after which each dark pattern is computed. 

\textit{Handling non-english text.} We extended the detection capabilities of \DD to handle multilingual cookie banners. Since consent banners appear in multiple languages, including EU-specific languages such as French and Polish, it is critical to ensure the system can process text from non-english sources. To address this, we added a language-detection algorithm using the \texttt{langdetect} library to identify the language of the extracted text. If the detected language is non-english, the text is translated into English using \texttt{argostranslate} before further processing. Additionally, the text is normalized by converting it to lowercase to ensure consistent pattern detection across different languages.

After completing the analysis, the output is saved in an SQLite database, with six tables as shown in the figure~\ref{fig:Overview}. The tool evaluates and detects 19 dark patterns, and allows researchers to extend it in the future if necessary.

In response to RQ1, \DD  demonstrated high detection accuracy, identifying the newly evolved dark patterns (DP11–DP19), with accuracy levels reaching up to 99\% for certain patterns across diverse datasets.
\subsection{Datasets}

We used two datasets for our system. For tool evaluation, we collected 500 random websites from the Top Tranco list. For our large-scale comprehensive measurement study, we compiled 14K websites (10K Top Tranco, 2K EU, 2K USA) crawled from January to October 2025, using input lists from CrUX, Majestic, Radar, and Umbrella.

\subsubsection{Manual human labelling dataset}
Cookie banners are relatively easy for humans to distinguish and to identify dark patterns. To create the ground-truth labeled dataset, the lead researcher manually labeled the dark patterns of consent banner from the top 500 websites. For efficiency, that snapshots of banner generated by \DD was used. 
In cases where multiple cookie banners appeared, the researcher analyzed all of them but selected the most prominently displayed one for validation. 
To evaluate the reliability of manual labeling, a second researcher independently reviewed a random 10\% subset (n = 50) of the dataset using the same snapshot collected by the \DD. And thus, 500 agreed ground-truth datasets were created, which were further used for tool's evaluation.


\subsubsection{Comprehensive measurement datasets} We constructed a dataset of 14,000 websites (10K from Top Tranco, 2K from the EU, and 2K from the USA) derived from the Tranco list~\cite{pochat2018tranco} of the most popular domains, which we then analyzed using our system. These datasets were collected between January 2025 and October 2025. To ensure accuracy and relevance, we initially downloaded the Top 1M list from Tranco. For the EU dataset, we customized the region-specific datasets from Tranco, and for the USA dataset, we filtered only .us domains to avoid overlap with the Top 10K Tranco .com websites. Finally, we removed all duplicates, resulting in a final dataset of 10K Top Tranco, 2K EU, and 2K USA websites, totaling 14,000 websites.



\begin{table}[t]
\centering
\caption{Evaluation Results for \DD (DP11 to DP19).}
\label{tab:evaluation_metrics}
\begin{tabular}{|c|c|c|c|c|}
\hline
\textbf{Dark Pattern} & \textbf{Accuracy} & \textbf{Precision} & \textbf{Recall} & \textbf{F1-Score} \\
\hline
DP11 & 0.9940 & 0.9959 & 0.9980 & 0.9969\\
DP12 & 0.9679 & 0.8614 & 0.9775 & 0.9158\\
DP13 & 0.9900 & 0.8889 & 0.9231 & 0.9057\\
DP14 & 0.9780 & 0.9593 & 0.9763 & 0.9677\\
DP15 & 0.9679 & 0.9717 &0.9639 & 0.9677\\
DP17 & 0.9238 & 0.8977 & 0.8876 & 0.8927\\
DP19 & 0.9739 & 0.9514 & 0.9778 & 0.9644\\
\hline
\end{tabular}
\end{table}

\subsection{Evaluation} 
\begin{figure*}[t]
    \centering
    \includegraphics[width=\linewidth]{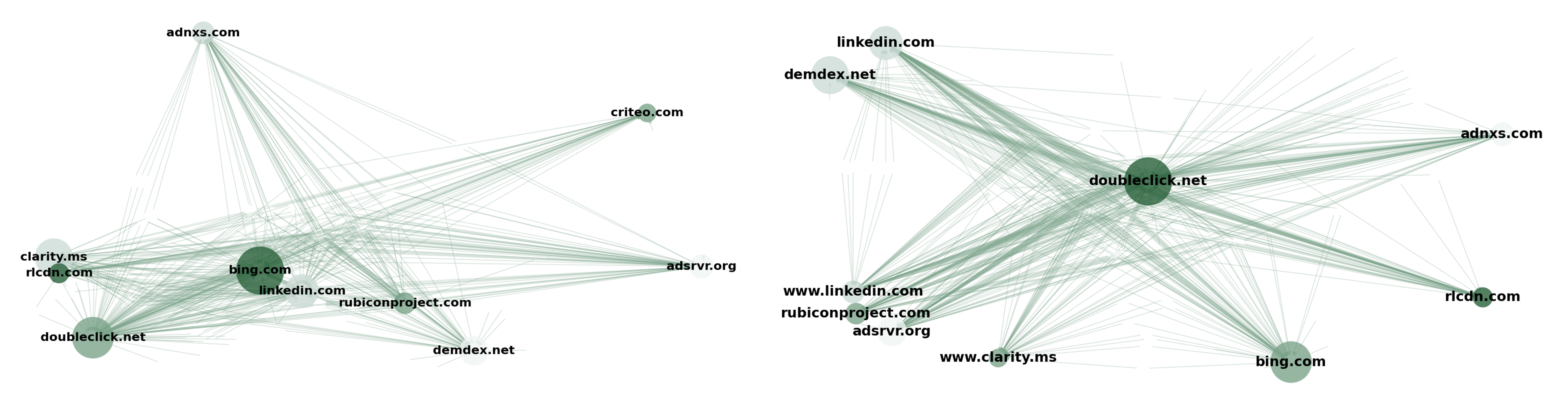} 
    \caption{Top 10 Third-Party tracker domains with Websites connected through edges (left: DP 18, right: DP 16).  This network graph illustrates the top 10 third-party tracker domains (e.g., doubleclick.net, linkedin.com, adnxs.com, and clarity.ms) and their connections to websites. Node size represents the frequency of website connections, while thicker edges indicate stronger associations. Across the network, doubleclick.net consistently emerges as the most central and dominant tracker domain.}
    \Description{Top 10 Third-Party tracker domains with Websites connected through edges (left: DP 18, right: DP 16).  This network graph illustrates the top 10 third-party tracker domains (e.g., doubleclick.net, linkedin.com, adnxs.com, and clarity.ms) and their connections to websites. Node size represents the frequency of website connections, while thicker edges indicate stronger associations. Across the network, doubleclick.net consistently emerges as the most central and dominant tracker domain.}
    \label{fig:NetEdge_DP16and_DP18}
\end{figure*}

To evaluate the effectiveness and reliability of the \DD tool, we compared its automated detection results with the manually labeled ground-truth dataset. The manually labeled ground-truth websites included sufficient samples for most dark patterns (DP11–DP19), with 490 instances for DP11 and 154 for DP12. Due to the low prevalence of DP13, we added news domains from the EU region, resulting in 26 instances. To ensure global representation, we randomly selected domains from the top 1 million Tranco list, covering regions like Germany (EU), Japan (Asia), and the USA (America). Our dataset shows 172 instances for DP14 and 249 for DP15. For DP16 and DP18, we ran the same domains through a simple cookie collector crawler and compared the results with those from \DD. The cookie properties, including the domain and other attributes, were identical in both \DD and simple crawler results, confirming the consistency and accuracy of the detected patterns.

Standard classification metrics—\textit{accuracy}, \textit{precision}, \textit{recall}, and \textit{F1-score}—were used to quantitatively assess the tool’s performance across different dark pattern categories. The evaluation outcomes for each dark pattern are summarized in Table~\ref{tab:evaluation_metrics}.

DP11 demonstrated near-perfect performance with an accuracy of 99.40\% and an F1 score of 0.9969, indicating excellent detection of this dark pattern. DP12 had lower precision (0.8614), suggesting the model often predicts True when it should predict False, but still maintained a high recall of 0.9775. DP17 showed the lowest performance with an accuracy of 92.38\% and a recall of 0.8876, indicating challenges in detection due to legal terms buried in various sub-pages. In contrast, DP19 performed well across all metrics, achieving an accuracy of 97.39\% and an F1 score of 0.9644, ranking as one of the model's better-performing patterns.

\begin{figure}[t]
    \centering
    \includegraphics[width=\linewidth]{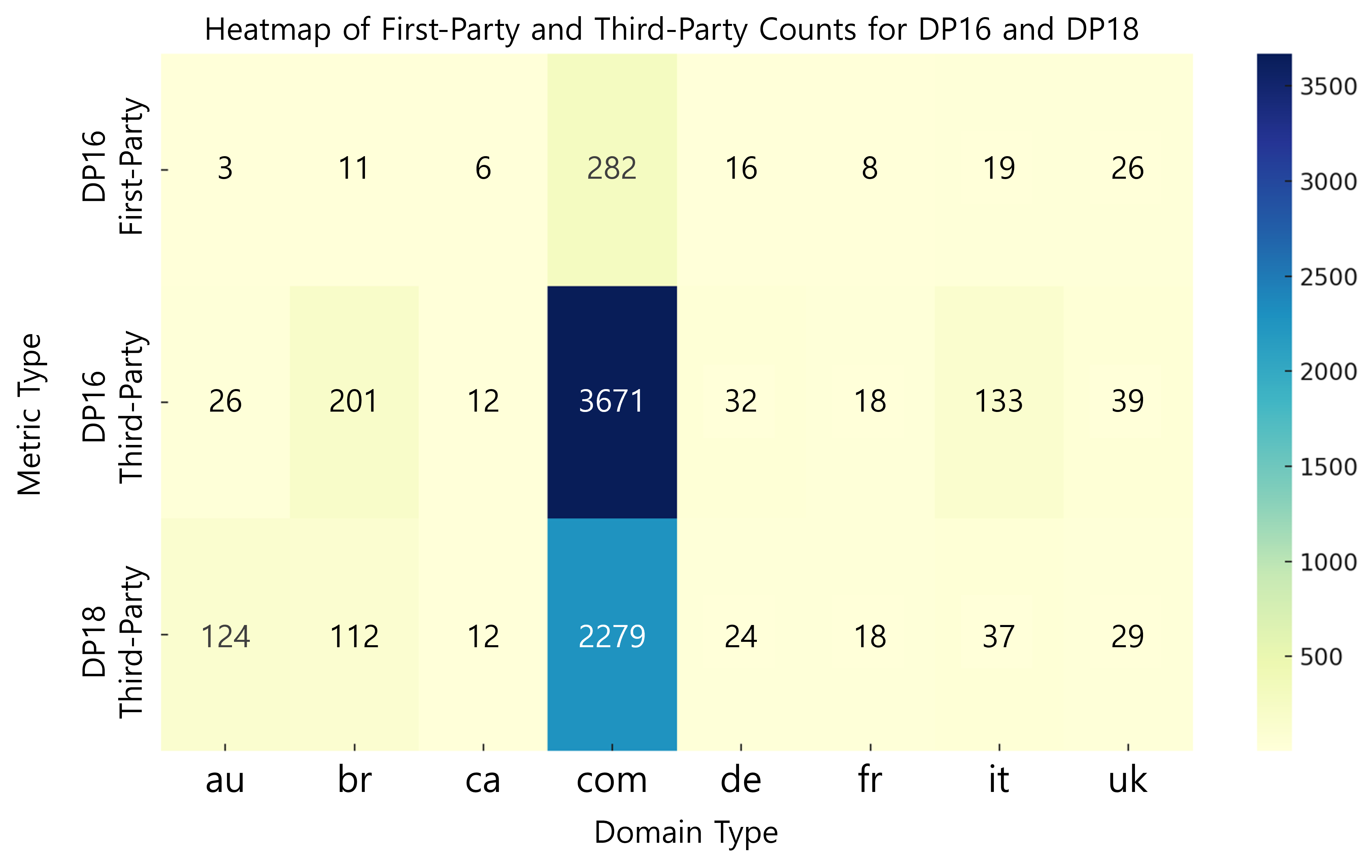} 
    \caption{The figure shows the counts of first-party cookies (DP16 First-Party) and third-party cookies (DP16 Third-Party) for different domains (au, br, ca, etc.). The darkest blue indicates the highest number of third-party cookies (3671 for ".com" in DP16), while light yellow indicates low counts. For DP18, only third-party domain counts are shown.} 
    \Description{The figure shows the counts of first-party cookies (DP16 First-Party) and third-party cookies (DP16 Third-Party) for different domains (au, br, ca, etc.). The darkest blue indicates the highest number of third-party cookies (3671 for ".com" in DP16), while light yellow indicates low counts. For DP18, only third-party domain counts are shown.} 
    \label{fig:Heatmap}
\end{figure}

\subsubsection{Evaluation for DP16 and DP18}
We evaluated cookie behavior by collecting cookies both pre-consent and post consent. Using the Disconnect filter list~\cite{Disconnect_Filter}, we identified known and popular tracking domains. 
DP16 and DP18 were examined by comparing the cookies set before and after user interaction with the consent banner. For DP16, \DD detects cookies set during the initial page load, before the user interacts with the banner. This is achieved by capturing the cookies during asynchronous loading using Selenium WebDriver, ensuring that all cookies, including first-party, third-party and ID-like cookies, are identified before consent is given. The tool waits for all page elements, including cookie-setting actions, to fully load before interacting with them. For instance, after clicking opt-out button, the tool waits for a 30-second delay to ensure that all cookies, including those set dynamically by third parties, are properly captured.  
For DP18, \DD uses dynamic cookie-state monitoring to confirm that no cookies are set post-rejection by comparing the cookies collected before and after user interaction. If cookies are still set despite rejection, DP18 is flagged.

Figure~\ref{fig:NetEdge_DP16and_DP18} reports the top 10 advertising trackers and their association with websites.
Figure~\ref{fig:Heatmap} shows the total distribution of First-party, third-party and trackers for DP16 and DP18.
For DP16, 346 domains set a total of 720 third-party cookies, with an average of 16.09 cookies per site; advertising cookies were the dominant tracker category. For DP18, when users opted out, we collected all cookies—first-party, third-party, and ID-like cookies—and found that 77\% of domains still allowed all three types of cookies, while 58\% of domains continued to set third-party cookies even after opt-out. In total, we observed 662 third-party domains for DP18, again dominated by advertising trackers, with an average of 10.85 cookies per site.

\section{Comprehensive Measurement Studies} To assess the global prevalence and behaviors of the detected dark patterns, we conducted large-scale measurements on 14K Top Tranco domains. We analyzed dark patterns, CMPs and their association with dark patterns, cookies and their properties, as well as other related factors across these regions.
\subsection{ Regional Comparison (EU vs US vs Top Tranco)}
Table~\ref{tab:Dataset} summarizes the regional datasets, detailing the number of websites loaded, consent dialogs extracted, and cookies collected out of the 14K sampled sites in each region.
\subsubsection{Prevalence of dark patterns over region (RQ2)} 
Table~\ref{tab:DP-Prevalence} presents a comparative analysis of dark pattern prevalence across these regions. When compared with the previous dark patterns identified by Kirkman et al. (DP1-DP10)~\cite{kirkman2023darkdialogs}, some domains no longer display certain dark patterns. For example, after facing substantial fines, many websites no longer show banners with only opt-in buttons (DP1), resulting in no prevalence for that pattern. In response to RQ2, our analysis reveals significant regional variation in the prevalence of newly evolved dark patterns (DP11–DP19) across the EU, US, and Tranco datasets. Such as, for the Tranco Top 10K list, DP11 has the highest prevalence (99\%), followed by DP16 at 68\%. DP13 is more prevalent in the EU region, while its presence in the USA and Tranco is minimal. This low prevalence in the USA and Tranco may be attributed to its legal conflicts. The distribution of DP19 is fairly consistent across all regions: 61\% in the EU, 63\% in the US, and 57\% in Tranco.

\begin{table}[t]
\centering
\caption{Consent dialogs data for different datasets.}
\resizebox{\columnwidth}{!}{  
\begin{tabular}{|p{3cm}|p{2.5cm}|p{2.5cm}|p{3cm}|}  
\hline
\rowcolor{gray!30} \textbf{Dataset} & \textbf{Loaded} & \textbf{Dialog} & \textbf{Cookies}\\
\hline
\textbf{2K EU } & 1871 & 838 & 38551 \\
\hline
\textbf{2K US} & 1459  & 84  & 11821  \\
\hline
\textbf{10K Tranco} & 8760  & 1038 & 174589  \\
\hline
\end{tabular}
}
\label{tab:Dataset}
\end{table}

\begin{table}[t]
\centering
\caption{Prevalence (\%) of dark patterns (DP1--DP19) across different regions.}
\begingroup
\footnotesize
\renewcommand{\arraystretch}{1.2}
\setlength{\tabcolsep}{2pt}
\begin{tabular}{
|>{\bfseries}m{2.3cm}|
>{\centering\arraybackslash}m{1.5cm}|
>{\centering\arraybackslash}m{1.5cm}|
>{\centering\arraybackslash}m{1.7cm}|
}
\hline
\rowcolor[HTML]{D9D9D9}
\textbf{Pattern} & \textbf{2K EU} & \textbf{2K US} & \textbf{10K Tranco} \\
\hline
Only Opt-In (DP1) & \cellcolor[HTML]{E6E6FA}0 (0\%) & \cellcolor[HTML]{E6E6FA}0 (0\%) & \cellcolor[HTML]{E6E6FA}0 (0\%) \\
Highlighted Opt-In (DP2) & \cellcolor[HTML]{FCE4D6}0 (0\%) & \cellcolor[HTML]{FCE4D6}0 (0\%) & \cellcolor[HTML]{FCE4D6}0 (0\%) \\
Obstruct Window (DP3) & \cellcolor[HTML]{FCE4D6}416 (49\%) & \cellcolor[HTML]{FCE4D6}34 (40\%) & \cellcolor[HTML]{FCE4D6}372 (35\%) \\
Complex Text (DP4) & \cellcolor[HTML]{FFF2CC}669 (79\%) & \cellcolor[HTML]{FFF2CC}28 (33\%) & \cellcolor[HTML]{FFF2CC}654 (63\%) \\
More Options (DP5) & \cellcolor[HTML]{FFF2CC}0 (0\%) & \cellcolor[HTML]{FFF2CC}0 (0\%) & \cellcolor[HTML]{FFF2CC}0 (0\%) \\
Ambiguous Close (DP6) & \cellcolor[HTML]{E6E6FA}0 (0\%) & \cellcolor[HTML]{E6E6FA}0 (0\%) & \cellcolor[HTML]{E6E6FA}0 (0\%) \\
Multiple Dialogs (DP7) & \cellcolor[HTML]{F2F2F2}91 (10\%) & \cellcolor[HTML]{F2F2F2}6 (7\%) & \cellcolor[HTML]{F2F2F2}78 (7\%) \\
Preference Slider (DP8) & \cellcolor[HTML]{F2F2F2}0 (0\%) & \cellcolor[HTML]{F2F2F2}0 (0\%) & \cellcolor[HTML]{F2F2F2}0 (0\%) \\
Close More Cookies (DP9) & \cellcolor[HTML]{F2F2F2}28 (3\%) & \cellcolor[HTML]{F2F2F2}2 (2\%) & \cellcolor[HTML]{F2F2F2}46 (4\%) \\
Opt-Out More Cookies (DP10) & \cellcolor[HTML]{D6EAD6}84 (10\%) & \cellcolor[HTML]{D6EAD6}3 (3\%) & \cellcolor[HTML]{D6EAD6}105 (10\%) \\
Cookie Info Display (DP11) & \cellcolor[HTML]{D6EAD6}837 (99\%) & \cellcolor[HTML]{D6EAD6}82 (97\%) & \cellcolor[HTML]{D6EAD6}1034 (99\%) \\
Purpose Info Display (DP12) & \cellcolor[HTML]{E6E6FA}428 (51\%) & \cellcolor[HTML]{E6E6FA}53 (63\%) & \cellcolor[HTML]{E6E6FA}406 (39\%) \\
Opt-Out Pricing (DP13) & \cellcolor[HTML]{F2F2F2}72 (8\%) & \cellcolor[HTML]{F2F2F2}3 (3\%) & \cellcolor[HTML]{F2F2F2}42 (4\%) \\
Revocation Impossible (DP14) & \cellcolor[HTML]{E6E6FA}358 (42\%) & \cellcolor[HTML]{E6E6FA}46 (54\%) & \cellcolor[HTML]{E6E6FA}422 (40\%) \\
Revocation Hard (DP15) & \cellcolor[HTML]{D6EAD6}275 (57\%) & \cellcolor[HTML]{D6EAD6}32 (38\%) & \cellcolor[HTML]{D6EAD6}432 (41\%) \\
Pre-Consent Cookies (DP16) & \cellcolor[HTML]{D6EAD6}375 (44\%) & \cellcolor[HTML]{D6EAD6}70 (83\%) & \cellcolor[HTML]{D6EAD6}712 (68\%) \\
Legal Ambiguity (DP17) & \cellcolor[HTML]{D6EAD6}170 (20\%) & \cellcolor[HTML]{D6EAD6}9 (10\%) & \cellcolor[HTML]{D6EAD6}258 (24\%) \\
Fake Opt-Out (DP18) & \cellcolor[HTML]{D6EAD6}196 (23\%) & \cellcolor[HTML]{D6EAD6}21 (25\%) & \cellcolor[HTML]{D6EAD6}253 (24\%) \\
Multi-Click Opt-Out (DP19) & \cellcolor[HTML]{D6EAD6}518 (61\%) & \cellcolor[HTML]{D6EAD6}53 (63\%) & \cellcolor[HTML]{D6EAD6}609 (57\%) \\
\hline
\end{tabular}
\label{tab:DP-Prevalence}
\endgroup
\end{table}

\begin{table}[t]
\centering
\caption{Data table showing the cookie properties for all cookies in the dataset(in percentage). }
\resizebox{\columnwidth}{!}{
\begin{tabular}{lccc} 
\hline
\rowcolor{gray!20} \textbf{Rank} & \textbf{2K (EU)} & \textbf{2K (US)} & \textbf{10K (Tranco)}  \\ 
\hline
\textbf{Total Cookies}                          & 38551            & 11821            & 174589                 \\
\textbf{Unique domains}                         & 3069             & 785              & 6922                   \\
\textbf{Secure (T/F)}                           & 76/24            & 67/33            & 75/25                  \\
\textbf{HttpOnly (T/F)}                         & 20/80            & 20/80            & 20/80                  \\
\textbf{Samesite (Null/Lax/None/Strict)}        & 22/10/65/1       & 29/10/57/2       & 20/10/69/1             \\
\textbf{Expires \textgreater{} 12 Months}       & 77               & 89               & 92                     \\
\hline
\end{tabular}
}
\label{tab:cookie-properties}
\end{table}

\subsubsection{Cookies and their concerning properties} 
Previous studies have highlighted concerns regarding the configuration of cookie properties, as improper settings can lead to critical security vulnerabilities~\cite{chen2023taking, squarcina2023cookie, khodayari2024great, singh2025crumbled}. Despite 7 years of GDPR, we have observed persistent security challenges arising from vulnerable cookie properties. As shown in Table~\ref{tab:cookie-properties}, 80\% of cookies are susceptible to potential Cross-Site Scripting (XSS) attacks, while 1\% in the EU, 2\% in the US, and 1\% globally are only best secure to Cross-Site Request Forgery (CSRF) attacks. Most of these cookies are persistent, living for over 12 months, thereby violating the GDPR's 12-month cookie lifecycle rule~\cite{Cookie_lifecycle}.

\textit{Association of insecure cookies and dark patterns.} We analyzed the cookie properties of domains exhibiting dark patterns and found that their cookies tend to be more insecure when these dark patterns are triggered. Specifically, we examined the cookie policies of domains with dark patterns and found that, in the EU, 86\% of insecure cookies (48\%) are vulnerable to XSS attacks, while only 8\% are protected against potential CSRF attacks. Alarmingly, over half of these cookies are long-lived (over 1 year). For example, the domain \url{www.snb.it}
features DP18, and all its cookies were insecure and vulnerable to XSS attacks, posing significant security and privacy risks, even with a consent banner.
In the USA, 93\% of insecure cookies (45\%) are vulnerable to XSS attacks, with only 6\% secured against CSRF attacks. Among these, 78\% are persistent cookies. For instance, \url{www.avenue.us}, which features DP15 alongside other DPs, had all cookies insecure and 70\% vulnerable to XSS attacks, with all cookies being persistent.
Similarly, in top Tranco domains, 90\% of insecure cookies (70\%) are vulnerable to XSS attacks, and only 3\% are safe against CSRF attacks. Moreover, 82\% of these cookies are long-lived. For example, \url{www.dailycaller.com}, which features DP16, DP18, and DP15, had no cookies secure against XSS, with 82\% insecure and 83\% persistent cookies, further compromising user security and privacy. To evaluate the effect of specific dark patterns, we checked how many cookies were set across the entire 14k dataset when DP15 was true. We found that websites with this pattern set 25\% more cookies on average, significantly increasing the exposure to security risks such as XSS attacks.

\subsubsection{Cookies and trackers} 
\begin{figure}[t]
    \centering
    \includegraphics[width=0.95\columnwidth]{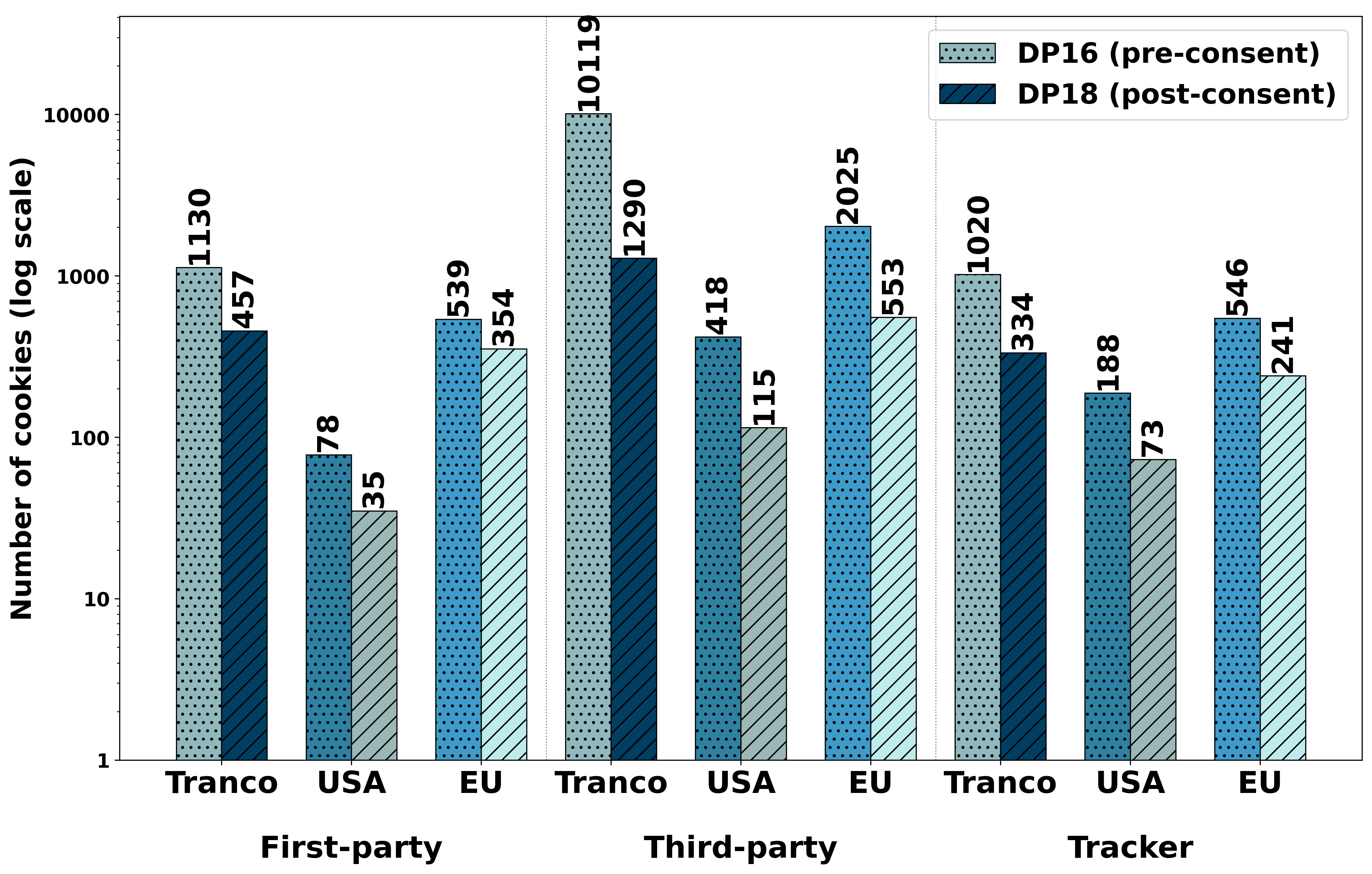}  
    \caption{Categorization of first-party, third-party, and tracker cookies collected from DP16 (Pre-Consent Cookies) and DP18 (Fake Opt-Out) across all datasets, presented as percentages for each region (EU, USA, Tranco).}
    \Description{Categorization of first-party, third-party, and tracker cookies collected from DP16 (Pre-Consent Cookies) and DP18 (Fake Opt-Out) across all datasets, presented as percentages for each region (EU, USA, Tranco).}
    \label{fig:dp16_18}
\end{figure}

We analyzed the cookies collected when DP16 and DP18 were set to True. The results are presented in Figure~\ref{fig:dp16_18} for all three regions. 
We categorized the cookies into first-party, third-party, and Trackers (using the Disconnect Filter list). 
This figure highlights the higher prevalence of post-consent cookies in the EU region compared to the USA and Tranco, with Tracker cookies exhibiting the most pronounced differences across the regions.

\begin{table*}[ht]
\centering
\caption{Mapping of prior work vs. dark patterns. Automated detection: {\Full}=automated dialog \& DP detection, {\Half}=automated either dialog extraction or DP detection, {\Empty}=manual. All dialog types: {\Full}=extracts all dialogs, {\Empty}=subset. Dark-pattern columns: {\Full}=measured, {\Half}=similar pattern measured, {\Empty}=mentioned only.}
\setlength{\tabcolsep}{4pt}
\renewcommand{\arraystretch}{1.1}
\resizebox{0.95\textwidth}{!}{%
\begin{tabular}{L{4.8cm} C{0.9cm}| C{0.60cm} C{0.60cm}| C{0.60cm} C{0.80cm}| C{0.80cm} C{0.80cm}| C{0.80cm} C{0.80cm}| C{0.80cm} C{0.80cm}| C{0.80cm} C{0.80cm}| C{0.80cm}C{0.80cm}C{0.80cm}}

 \toprule
& \vheader{Year}
& \vheader{\# Domains}
& \vheader{\# Dialogs}
& \vheader{All dialogs}
& \vheader{Cookie Classification}
& \vheader{Automated Detection}
& \vheader{Regional Comparision}
& \vheader{CookieInfoDisplay}
& \vheader{PurposeInfoDisplay}
& \vheader{OptOutPricing}
& \vheader{ConsentRevocationPossible}
& \vheader{RevocationHard}
& \vheader{PreConsentCookies}
& \vheader{LegalAmbiguity}
& \vheader{FakeOptOut}
& \vheader{MultiClick OptOut}\\
\midrule

Nouwens et al.~\cite{nouwens2020dark}& 2020 & 10k & 680  & \Full   & \Empty  & \Full & \Half  & \Empty  & \Empty & \Empty & \Empty & \Empty  & \Empty & \Empty &\Empty  &\Full \\

Matte et al.~\cite{matte2020cookie} &2020 & 28.2k & 1.4k  & \Full  & \Empty  & \Full  & \Half  & \Empty & \Empty & \Empty  & \Empty & \Empty  & \Full & \Empty  &\Full &\Full\\

Kampanos \& Shahandashti~\cite{kampanos2021accept} & 2021 & 17.7k & 7.5k & \Full  & \Full  & \Full  & \Half  & \Empty & \Empty & \Empty  & \Empty & \Empty  & \Empty & \Empty  &\Empty &\Full\\

Hils et al.~\cite{hils2020measuring} & 2021 & 40.2m & 414k  & \Full  & \Full  & \Full  & \Full  & \Full  & \Empty & \Full  & \Empty & \Full   & \Empty & \Empty  &\Empty &\Empty\\

Habib et al.\cite{habib2022okay} & 2021 & 500   & 255   & \Half  & \Empty & \Half  & \Empty & \Empty & \Empty & \Half  & \Empty  & \Empty & \Empty & \Empty  &\Empty &\Empty\\

Bollinger et al.~\cite{bollinger2022automating} & 2022 & 29.4k & --    & \Full  & \Full  & \Full  & \Empty  & \Empty  & \Half & \Empty  & \Empty  & \Empty & \Full & \Empty  &\Full &\Empty\\

Kirkman et al.~\cite{kirkman2023darkdialogs}    & 2023 & 11k   & 2k    & \Full & \Empty & \Full & \Half  & \Empty  & \Empty & \Empty & \Empty & \Empty & \Half & \Empty  &\Half &\Half\\

bouhoula et al.~\cite{bouhoula2024automated}    & 2024 & 97k   & --    & \Full & \Half & \Full & \Full  & \Empty & \Full & \Empty & \Empty & \Empty &  \Half &\Empty  &\Full &\Empty \\

Kachrela et al.~\cite{kancherla2025johnny}      & 2025 & 200   & --   & \Full & \Half & \Half & \Empty & \Empty & \Empty & \Empty & \Half & \Full & \Empty & \Empty &\Empty &\Empty \\
\textbf{Us (this work)}  & \textbf{2025} & \textbf{14k} & \textbf{2k} & \textbf{\Full} & \textbf{\Full} & \textbf{\Full} & \textbf{\Full} & \textbf{\Full} & \textbf{\Full} &\textbf{\Full} & \textbf{\Full}  & \textbf{\Full} & \textbf{\Full}& \textbf{\Full} & \textbf{\Full}& \textbf{\Full}\\
 \bottomrule
\end{tabular}%
}
\label{tab:related_work}
\end{table*}


\subsubsection{CMP vs Non-CMP patterns (RQ2)}
Table~\ref{tab:CMP} presents the prevalence of variou CMP across the three regions. OneTrust and Cookiebot are the most prevalent across all regions, with OneTrust being dominant in region. The data also show that Quantcast, TrustArc, and Didomi are prominent in the EU, whereas platforms such as Termly and DataPrivacyManager are less common. This distribution underscores regional variations in CMP usage, with certain platforms being more prevalent in specific regions. Additionally, we observed the presence of a newly launched CMP, `Google,' in the Tranco dataset. 

\begin{figure}[t]
    \centering
    \includegraphics[width=1\columnwidth]{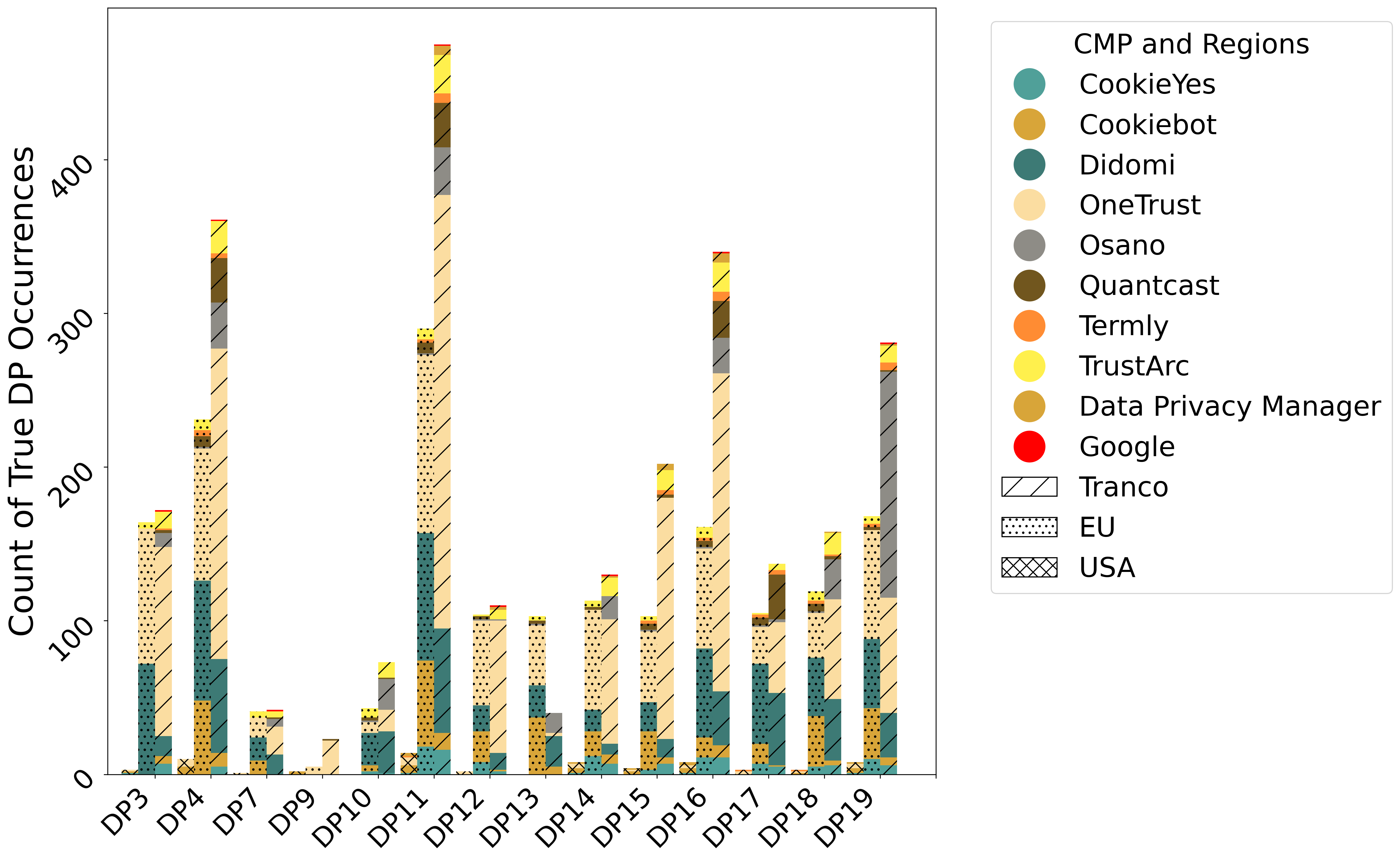}  
    \caption{Prevalence of dark patterns associated with all CMP for each region (EU, USA, Tranco).}
    \Description{Prevalence of dark patterns associated with all CMP for each region (EU, USA, Tranco).}
    \label{fig:CMP_DP}
\end{figure}

We analyzed the dark patterns associated with specific CMPs, revealing significant regional and platform-based variations. As shown in Figure~\ref{fig:CMP_DP}, Didomi and OneTrust are linked to more manipulative patterns like HighlightedOptIn and ObstructsWindow, particularly in the EU and Tranco regions. In contrast, USA CMPs like Termly and Quantcast exhibit fewer dark patterns, likely due to less stringent regulations. This suggests CMPs tailor their designs to regional compliance and user behavior strategies.

\begin{table}[t]
\centering
\caption{Prevalence of CMP across regions, with the numbers representing all datasets.}
\small
\begin{tabular}{|p{3cm}|p{1.2cm}|p{1.2cm}|p{1.1cm}|}  
\hline
\rowcolor{gray!20}\textbf{CMP Platform} & \textbf{EU} & \textbf{USA} & \textbf{Tranco} \\ \hline
CookieYes& 18 & 9 & 41 \\
Cookiebot& 60 & 9 & 27 \\
Didomi & 84 & 1 & 122 \\
OneTrust & 116 & 47 & 641 \\
Quantcast & 7 & 0 & 42 \\
TrustArc & 7 & 3 & 49 \\
Osano & 1 & 4 & 64 \\
Termly & 2 & 2 & 9 \\
DataPrivacyManager & 0 & 6 & 16 \\
Google & 0 & 0 & 1\\
\hline
\end{tabular}
\label{tab:CMP}
\end{table}

\subsection{Clickables} 
\begin{figure}[H]
    \centering
    \includegraphics[width=1\columnwidth]{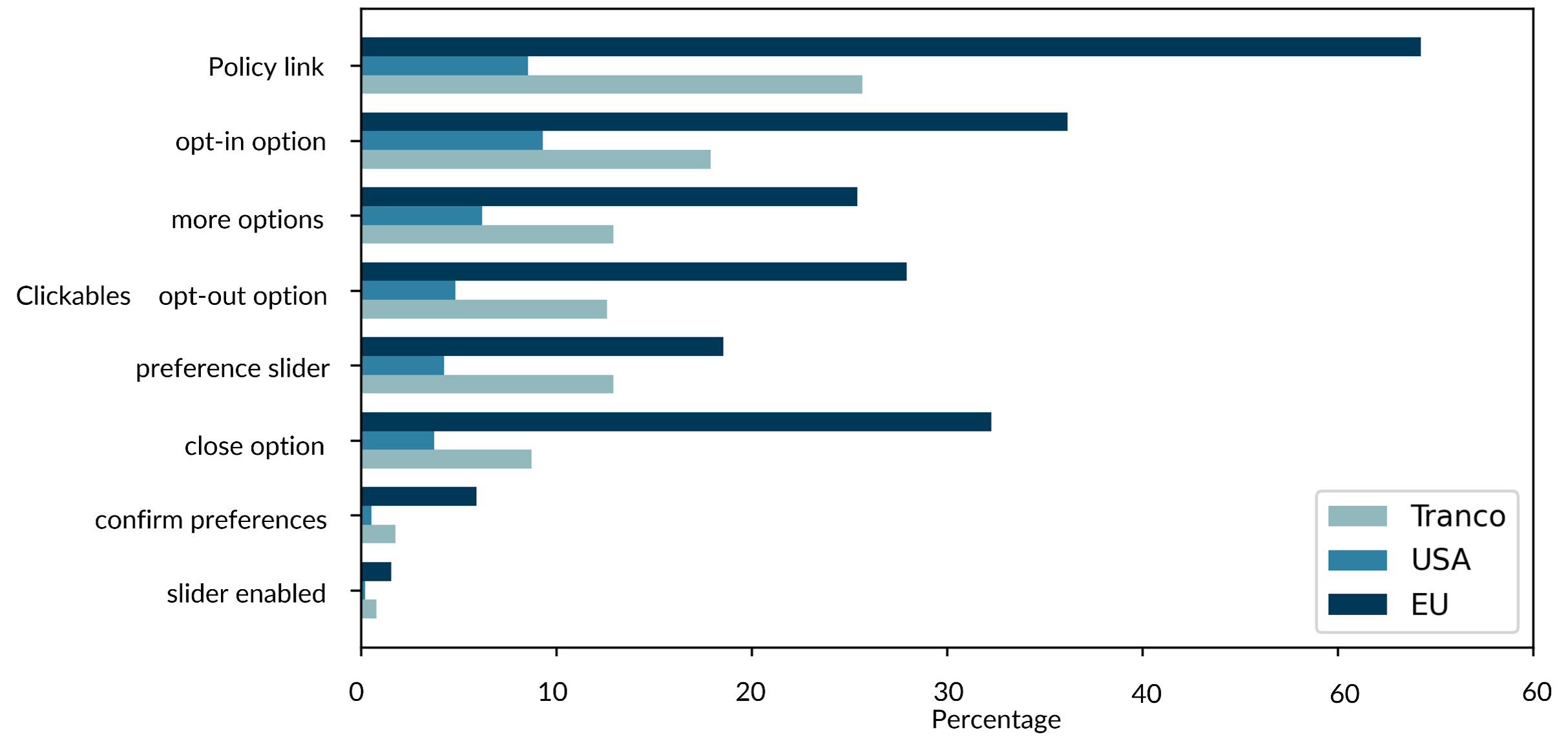}  
    \caption{Figure showing the percentage of websites with at least one occurrence of each clickable type on cookie banner.}
    \Description{Figure showing the percentage of websites with at least one occurrence of each clickable type on cookie banner.}
    \label{fig:clickables}
\end{figure}

Cookie banners feature various clickable items for user interaction. Figure~\ref{fig:clickables} presents the common clickable buttons found across all regions. As shown, most banners in the EU region include a policy link, which is also the most common across other regions. Although the appearance of preference sliders is minimal, their presence can pose privacy risks with even the slightest negligence or misuse. 

\section{Discussion}
Our findings reveal widespread non-compliance, particularly regarding consent revocation barriers and covert tracking, underscoring the need for more effective regulatory oversight. 

Despite the explicit requirement that websites implement consent revocation mechanisms as outlined in Article 7 (3)~\cite{Article_7_3}, our findings show that while 58\% of EU domains, 46\% of US domains, and 60\% of global Tranco domains have incorporated these mechanisms, they still fail to comply with the regulation that revocation should be as easy as giving consent (Article 7 (2))~\cite{Article7_2}. Therefore, 57\% of domains in the EU have a revocation mechanism that is difficult for users to opt into.
Cookie banners have continuously exhibited dark patterns over time, with domains making efforts to update their banners periodically in order to fully comply with regulations (see Appendix Figure~\ref{fig:Evolved_banner1_old} and ~\ref{fig:Evolved_banner1_new}). However, despite these efforts (see Appendix Table~\ref{tab:Legal_Mapping}), domains not only introduce new dark patterns but also continue to deploy insecure cookies.
We analyzed domains where consent revocation was classified as either ``hard'' or ``not hard'' and compared the attributes of cookies collected from these domains. Our findings reveal that, even when DP15 is absent, users remain vulnerable to security risks due to improper cookie settings for attributes like `HttpOnly,' `Secure,' and `Samesite.' This underscores that eliminating dark patterns does not fully mitigate security risks if cookie configurations are not properly managed.

Our study demonstrates that cookie consent mechanisms, especially those featuring dark patterns like ``Fake Opt-Out (DP18)'' and ``Pre-Consent Cookies (DP16),'' significantly compromise security and privacy. These mechanisms often result in the setting of tracker cookies without user consent and even after consent is given. 

Additionally, these cookies frequently possess insecure properties, exacerbating the associated security and privacy risks. 

While the above findings address significant regulatory issues, we also identify the importance of transparency in banner. For instance, DP11, which calls for a clear, plain-language definition of cookies—is essential for ensuring informed consent, particularly for users who may not be familiar with technical cookie terminology. Although not explicitly required in every instance, including a clear definition in cookie banners, as seen in widely used CMPs such as OneTrust (Figure 2), aligns with the GDPR's requirement for transparency (Article 12 (1))~\cite{GDPR_Article_12} and ensures that users understand what cookies are and how they function.

\section{Limitations and Future Work}

Our study has several limitations that frame the scope of our findings. First, although \DD is designed to be CMP-agnostic, the rendering of banner designs can vary significantly across devices. Consequently, differences in the DOM structure between mobile and desktop views may impact detection accuracy in specific edge cases. Furthermore, while we utilize Selenium for dynamic interaction, highly obfuscated implementations or adversarial changes to the DOM structure designed to evade detection remain a persistent challenge in automated web measurement.

Second, regarding security implications, we focused on identifying vulnerable cookie configurations (e.g., missing \texttt{HttpOnly} or \texttt{Secure} flags) and their correlation with dark patterns. We did not generate active exploits (e.g., XSS payloads) to validate compromise. This decision was made to ethically limit the study to measuring the expanded \textit{attack surface} and identifying poor security hygiene, rather than confirming individual breaches on live websites.

Third, our detection logic relies on deterministic heuristics and defined lexicons. While this approach achieves high precision for known patterns (DP1--DP19), it may not immediately generalize to entirely novel, undefined manipulation strategies without rule updates. However, the system is designed to be extensible, allowing for the integration of emerging patterns through manual validation and rule refinement.

Finally, our dataset is biased toward top-ranked Tranco domains and major European languages. Long-tail websites or those in under-represented languages may exhibit different non-compliance patterns that our current scope might miss. Future work will extend this analysis to a broader range of non-English and non-EU websites to enhance the generalizability of our findings.

\section{Related Work} 
User consent collection, established with the web, gained prominence with the EU’s 2009 ePrivacy Directive and the GDPR, which set strict rules for personal data collection. This led to studies on CMPs and their impact on consent interfaces and user choices~\cite{hils2020measuring}. 
Table~\ref{tab:related_work} compares prior research on cookie consent banners and dark patterns, summarizing dataset size, automation level, and addressed dark pattern categories to highlight differences in coverage and methodology.
Similar to our findings, previous studies~\cite{nouwens2020dark} have highlighted how dark patterns, such as pre-ticked boxes and difficult rejection options, undermine user consent in CMPs, violating GDPR compliance.

Matte et al.~\cite{matte2020cookie} also found that CMP banners disrespect user consent choices, provide no option to opt-out, and prompt consent before the user has a chance to make an informed decision. In contrast, we found that these dark patterns persist even after several years, accompanied by vulnerable cookie policies.
Kampanos \& Shahandashti~\cite{kampanos2021accept} similarly found in their large-scale study across the UK and Greece that, despite regulations, only half of the domains with third-party cookies display cookie banners. Among those, most use nudge designs that make consent more difficult, similar to our findings, where users must click multiple times to reject cookies.
Despite the doubling of CMP involvement on websites, nudge designs persist~\cite{habib2022okay,hils2020measuring}, introducing usability issues that compromise user privacy. In our study, we extended this analysis by revealing that, in addition to these privacy concerns, these banners often contain security vulnerabilities as well, further exacerbating the risks for users.
Bolinger et al.~\cite{bollinger2022automating} classified cookies on 30K websites and found that well-known cookies often had incorrect purposes defined and were forcefully set, either before or after negative consent.
In previous work~\cite{kirkman2023darkdialogs}, all 10 dark patterns were detected and identified; however, in light of the evolution of dark patterns, these patterns are now considered outdated. To address this, we have identified 9 novel dark patterns. Similarly, previous studies overlook dark patterns related to cookie behavior and their characteristics~\cite{bouhoula2024automated}. In contrast, our work provides an in-depth analysis to better understand their behavior and predict the security challenges associated with them.
The challenges of revocation and cookie paywalls have evolved into dark patterns, and many researchers are actively studying these critical issues. For instance, an interesting study by ~\cite{kancherla2025johnny} has been conducted in this context. However, to date, no study has examined these dark patterns through large-scale measurement studies. Our work provides a comparative analysis of these evolved dark patterns using a global dataset, depicting their prevalence.

\section{Ethical Considerations} 
In conducting this research, the \DD adhered to ethical guidelines to ensure compliance with web scraping best practices. Specifically, the tool respected the \texttt{robots.txt} directives and excluded any websites that explicitly prohibited crawling or data collection.  Additionally, we ensure that the crawling and analysis do not negatively impact the performance of the websites being analyzed.
\section{Conclusion}
In this work, we introduced \DD, an automated system that detects nine evolved dark patterns by combining text analysis, visual cues, and cookie-state verification in a CMP-agnostic manner. Our findings show that improper cookie policies combined with dark patterns can not only undermine users' privacy choices but also introduce potential security risks, such as XSS and CSRF attacks.

Our large-scale measurement shows that dark patterns remain widespread, undermining users’ ability to make informed privacy decisions. By explicitly mapping these patterns to GDPR and EDPB guidelines, \DD provides not only a measurement framework but also a practical compliance lens for regulators and policymakers.
By bridging automated detection with regulatory oversight, \DD contributes to a more transparent and user-respecting web ecosystem.
\bibliographystyle{plainnat}
\bibliography{References1}
\appendix
\section{Lexicons}
 \label{Appendix:Lexicons}
Table~\ref{table:Lexicons}, we present the lexicons used in \DD. The table includes the lexicons for DP11, DP12, DP13, and DP17.
\begin{table*}[ht!]
\centering
\caption{This table presents the complete set of lexical indicators used to detect five categories of dark patterns:  CookieInfoDisplay (DP11), PurposeInfoDisplay (DP12), OptOutPricing (DP13), and LegalAmbiguity (DP17).}
\Description{This table presents the complete set of lexical indicators used to detect five categories of dark patterns:  CookieInfoDisplay (DP11), PurposeInfoDisplay (DP12), OptOutPricing (DP13), and LegalAmbiguity (DP17).}
\label{table:Lexicons}
\resizebox{0.95\linewidth}{!}{
\begin{tblr}{
  colspec={p{2cm} p{2cm} p{14cm}},
  cell{3}{1} = {r=4}{},
  cell{7}{1} = {r=2}{},
  cell{9}{1} = {r=2}{},
  hline{1-3,7,9,11} = {-}{},
  hline{4-6,8,10} = {2-3}{},
}
\textbf{Dark Patterns}                           & \textbf{Category}            & \textbf{Lexicon}                                                                                                                                                                                                                                                                                                                                                                                                                                                                                                                                                                                                                                                                                                                                                                                                                                                                                                                                                                                               \\
{\textbf{CookieInfo\\Display }\\\textbf{(DP11)}}   &             -                 & "cookies are small text files," "cookies are small files," "cookies (small text files," "cookies contain," "cookies hold," ~"cookies identify," "cookies track," "Cookies and other tools store or retrieve personal data," "script (e.g. cookies)," ~"script such as cookies," "small files called cookies," "A cookie is a small text file,"                                                                                                                                                                                                                                                                                                                                                                                                                                                                                                                                                                                                                                                                    \\
{\textbf{PurposeInfo\\Display (DP12)}} & Analytics                    & {"for analytics," "to analyze website traffic," "to analyze user behavior," "for collecting statistical data," ~"for monitoring site performance," "to analyze traffic," "to assess the performance of ads," \\"to access personal data for analytics," "analyse site traffic," "analze our traffic," "for statistical purposes," ~"for measuring performance," "measure site performance," "for A/B testing,"}                                                                                                                                                                                                                                                                                                                                                                                                                                                                                                                                                                                                   \\
                                                 & Advertising / Targeting      & "for marketing," "for advertising," "to deliver targeted ads," "for better targeting," "for retargeting," "for affiliate marketing," ~"for improving ads efficiency," "for customizing advertisements," "to serve advertising," "profiling," "personalized advertising," ~"interest-based advertising," "relevant personalized advertisements," "third party advertising purposes," ~"to access personal data for advertising," "advertising cookies," "to show relevant ads,"
\\

                                                 & Personalization / Experience & "to personalize content," "to personalize," "for personalization," "to remember your preferences," "remember their preferences," ~~"personalization of communication," "to improve user experience," "to help improve your experience," "to improve your experience," "to provide you with the best experience," "to tailor content and ads," "to enhance site content," "store and access information," "to provide content recommendations," "to recommend products," "cookies save your preferences," "cookies help websites remember," "cookies allow websites," "personalize our site," "to enhance,"                                                                                                                                                                                                                                                                                                                                                                                                     \\
                                                 & Functionality / Performance  & "for functional purposes," "for functionality," "for site functionality," "improve site functionality," "to optimize our website," "to improve site performance," "for site security," "Authentication and security," "to process device information," "process personal data," "to access personal data for social engineering," "to maintain session state," "for session management," "to enable website features," "to enable features like chat," "for managing your shopping cart," "to facilitate payment processing," "for user authentication," "to support customer support," "to provide social media features," "for better site navigation," "proper operation of the website," "to deliver and enhance the quality of services," "to provide better services," "creating custom content profile," "for geo-targeting," "to gather demographic information," "to track usage," "to track affiliate links," "performance cookies," "functionality cookies," "optimizing," "remember your settings," \\
                                                 
{\textbf{OptOutPricing}\\\textbf{(DP13)}}        & {Pay terms}                 & "subscribe," "subscription," "pay," "price," "pricing," "per month," "per week," "per year," "trial," "ad-free," "premium," "€," "\$" \\ 

                                                 & {Consent terms}             & "cookie," "cookies," "consent," "tracking," "reject," "refuse," "deny," "opt-out," "without ads," "without tracking," "without advertising," "ad-free," "with ads," "with tracking," "purchase"                                                                                                                                                                                                                                                                                                                                                                                                                                                                                                                                                                                                                                                                                                                                                                                                                \\
{\textbf{Legal\\Ambiguity}\\\textbf{(DP17)}}       & {Legal basis\\terms}         & "GDPR Article 6," "legitimate interest (GDPR)," "contract performance for cookies," "legitimate interest (data protection regulation)," "cookie processing under GDPR," "ccpa consent," "edpb guidelines," "processing based on consent," "GDPR," "gdpr," "edpb," "EDPB," "eprivacy," "cookie tracking consent," "article 6," "article," "ccpa," "legal basis for cookies," "legitimate interest (gdpr)," "consent to cookie use," "data processing consent under GDPR," "CCPA," "California Consumer Privacy Act (CCPA)," "article 7," "Article 6 (1)," "CCPA consent," "GDPR Article 6 (1)(a)," "california notice at collection," "Addiotional disclosures for california residents," "Consent to Cookie Use (GDPR)," "ePrivacy Regulation," "ePrivacy compliant," "EDPB Guidelines," "EDPB Recommendations," "European Data Protection Board,"                                                                                                                                                              \\
                                                 & {Activity terms}            & "process personal data," "collect personal data," "track personal data," "use personal data," "store personal data," "access personal data," "retrieve personal data," "share personal data," "transfer personal data," "analyze personal data," "manage personal data," "process cookies," "track cookies," "store cookies," "track user behavior," "personalize content," "serve personalized ads," "advertise products," "recommend products," "measure website usage," "analyze traffic," "monitor traffic," "optimize user experience," "monitor website performance," "content personalization," "geo-targeting," "personalized content," "tracking technologies," "user tracking," "user profiling," "process your data," "targeted advertising profiling," "process user's data," "advertising profiling," "consumer profiling," "behavioral targeting,"                                                                                                                                                
\end{tblr}
}
\label{tab:lexicons}
\end{table*}

\section{Legacy and Evolved Cookie Consent Banners} 
\begin{figure}[H]
    \centering
    \includegraphics[width=0.95\linewidth]{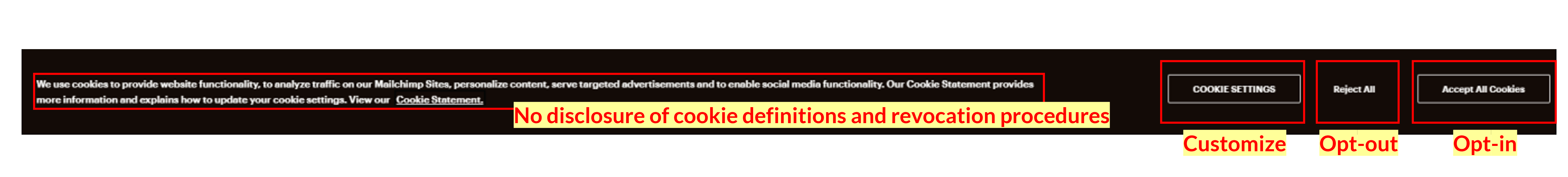}
    \caption{The legacy cookie banner, as discussed in the article~\cite{kirkman2023darkdialogs}, represents older designs with limited user control and clarity.}
    \Description{The legacy cookie banner, as discussed in the article~\cite{kirkman2023darkdialogs}, represents older designs with limited user control and clarity.}
    \label{fig:Evolved_banner1_old}
\end{figure}

\begin{figure}[H]
    \centering
    \includegraphics[width=\linewidth]{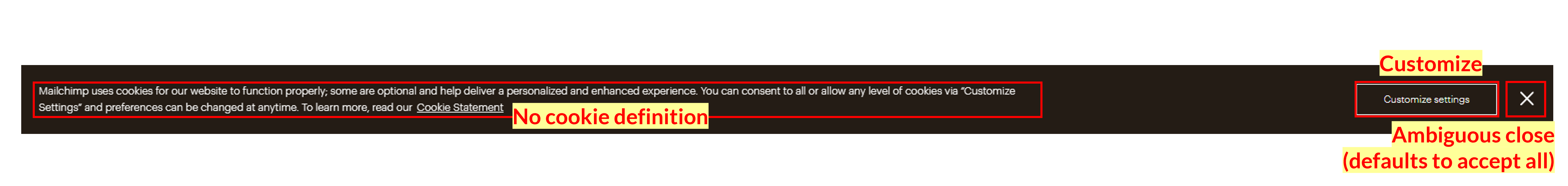}
    \caption{The evolved cookie banner currently on the source site includes dark patterns, such as ambiguous close options and DP19, which have emerged in more recent designs.}
    \Description{The evolved cookie banner currently on the source site includes dark patterns, such as ambiguous close options and DP19, which have emerged in more recent designs.}
    \label{fig:Evolved_banner1_new}
\end{figure}

\begin{figure}[H]
    \centering
    \includegraphics[width=0.95\linewidth]{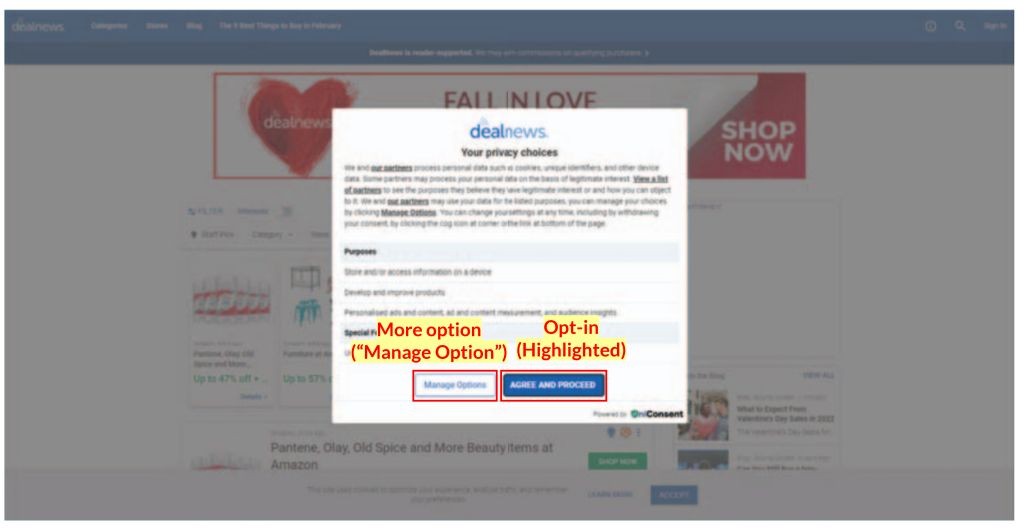}
    \caption{The legacy cookie banner, as discussed in the article~\cite{kirkman2023darkdialogs}, represents the more options and DP2.}
    \Description{The legacy cookie banner, as discussed in the article~\cite{kirkman2023darkdialogs}, represents the more options and DP2.}
    \label{fig:Evolved_banner2_old}
\end{figure}

\begin{figure}[H]
    \centering
    \includegraphics[width=\linewidth]{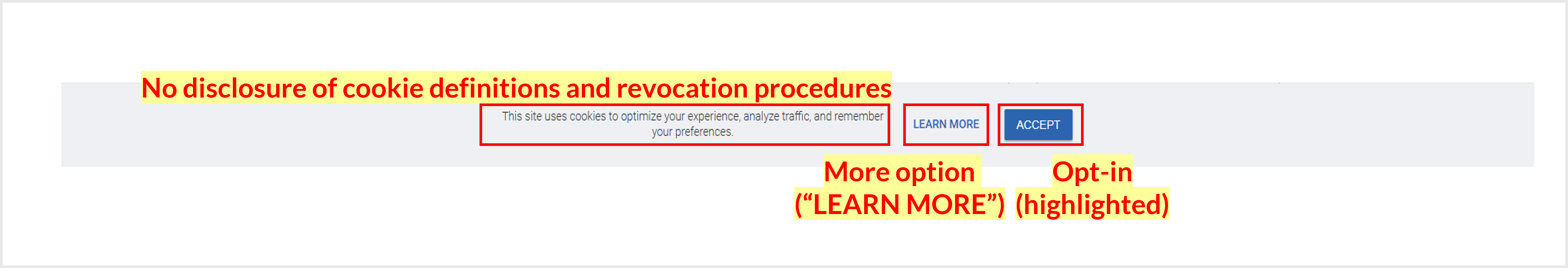}
    \caption{The evolved cookie banner currently on the source site includes dark patterns, DP1, but the information is reduced.}
    \Description{The evolved cookie banner currently on the source site includes dark patterns, DP1, but the information is reduced.}
    \label{fig:Evolved_banner2_new}
\end{figure}

\begin{figure}[H]
    \centering
    \includegraphics[width=0.95\linewidth]{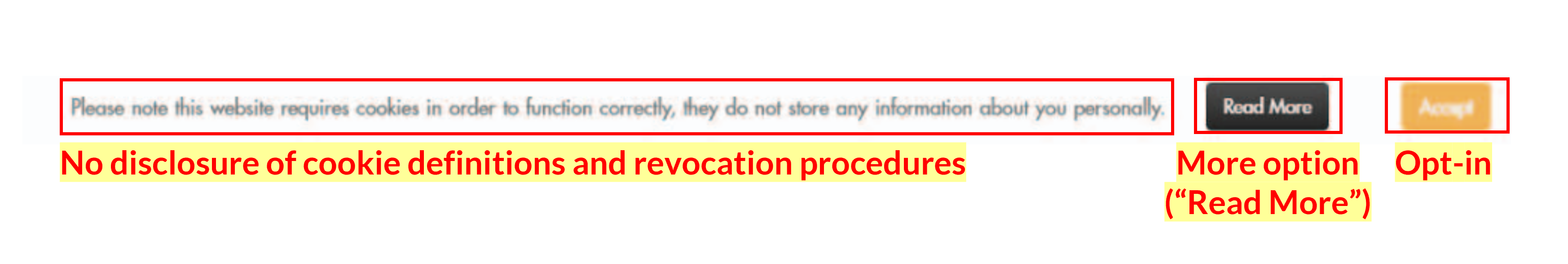}
    \caption{The legacy cookie banner, as discussed in the article~\cite{kirkman2023darkdialogs}, represents the more options and DP2.}
    \Description{The legacy cookie banner, as discussed in the article~\cite{kirkman2023darkdialogs}, represents the more options and DP2.}
    \label{fig:Evolved_banner3_old}
\end{figure}

\begin{figure}[H]
    \centering
    \includegraphics[width=0.80\linewidth]{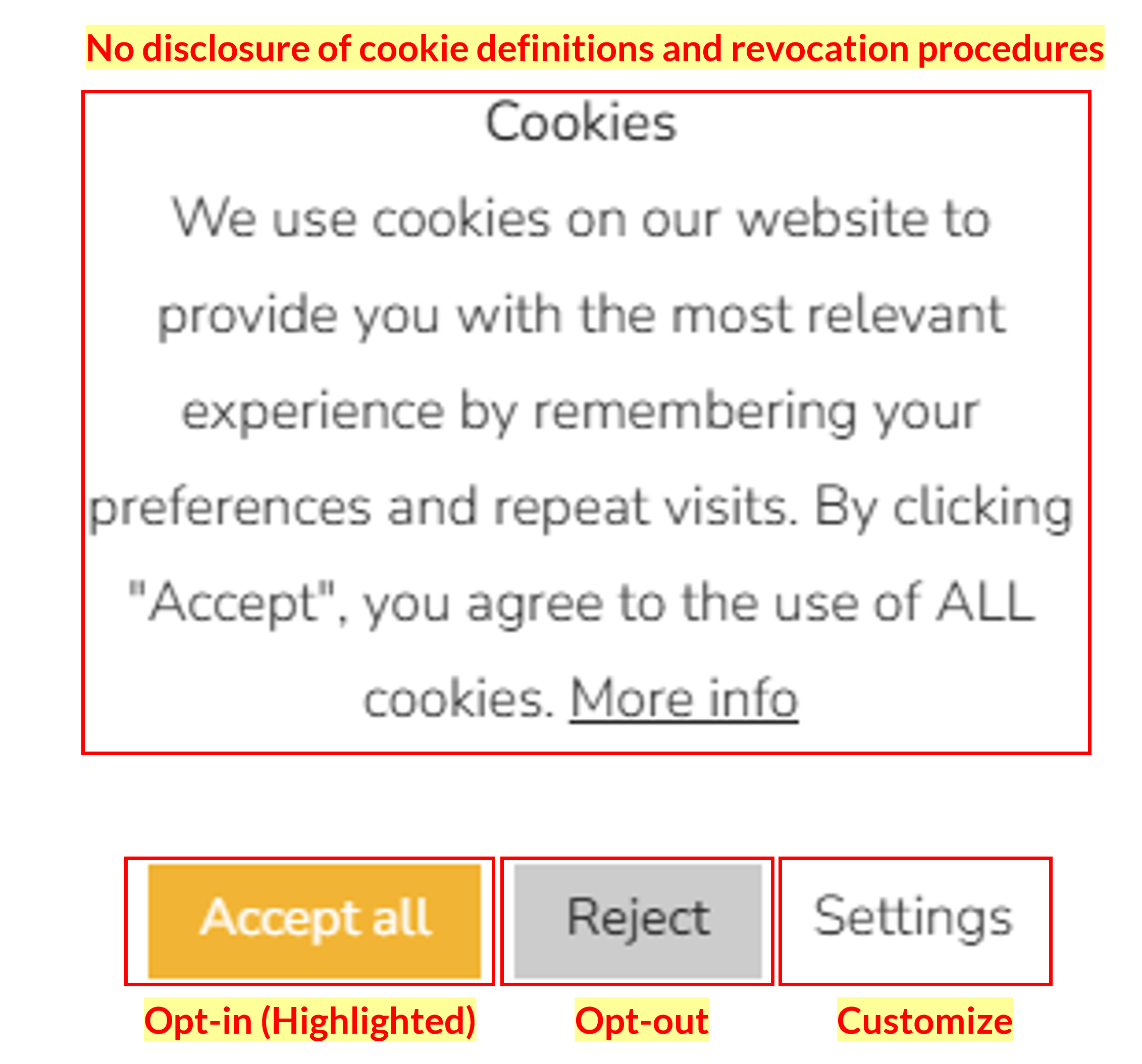}
    \caption{The evolved cookie banner currently on the source site includes dark patterns specifically, DP11 and DP12.}
     \Description{The evolved cookie banner currently on the source site includes dark patterns specifically, DP11 and DP12.}
    \label{fig:Evolved_banner3_new}
\end{figure}

\begin{figure}[H]
    \centering
    \includegraphics[width=0.95\linewidth]{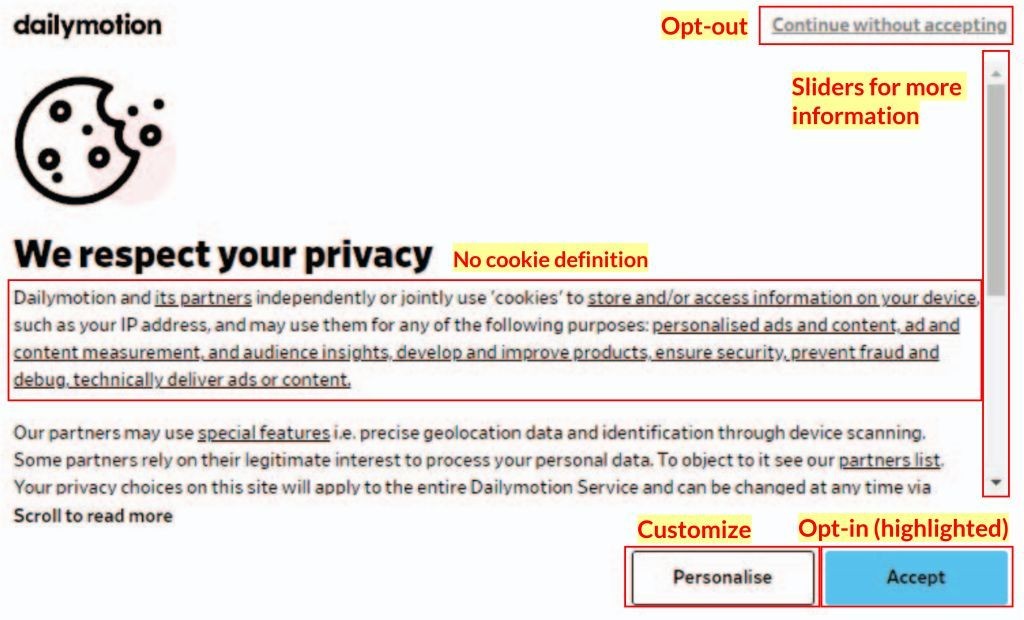}
    \caption{The legacy cookie banner, as discussed in the article~\cite{kirkman2023darkdialogs}, includes detailed opt-out options and additional information links.}
    \Description{The legacy cookie banner, as discussed in the article~\cite{kirkman2023darkdialogs}, includes detailed opt-out options and additional information links.}
    \label{fig:Evolved_banner4_old}
\end{figure}

\begin{figure}[H]
    \centering
    \includegraphics[width=\linewidth]{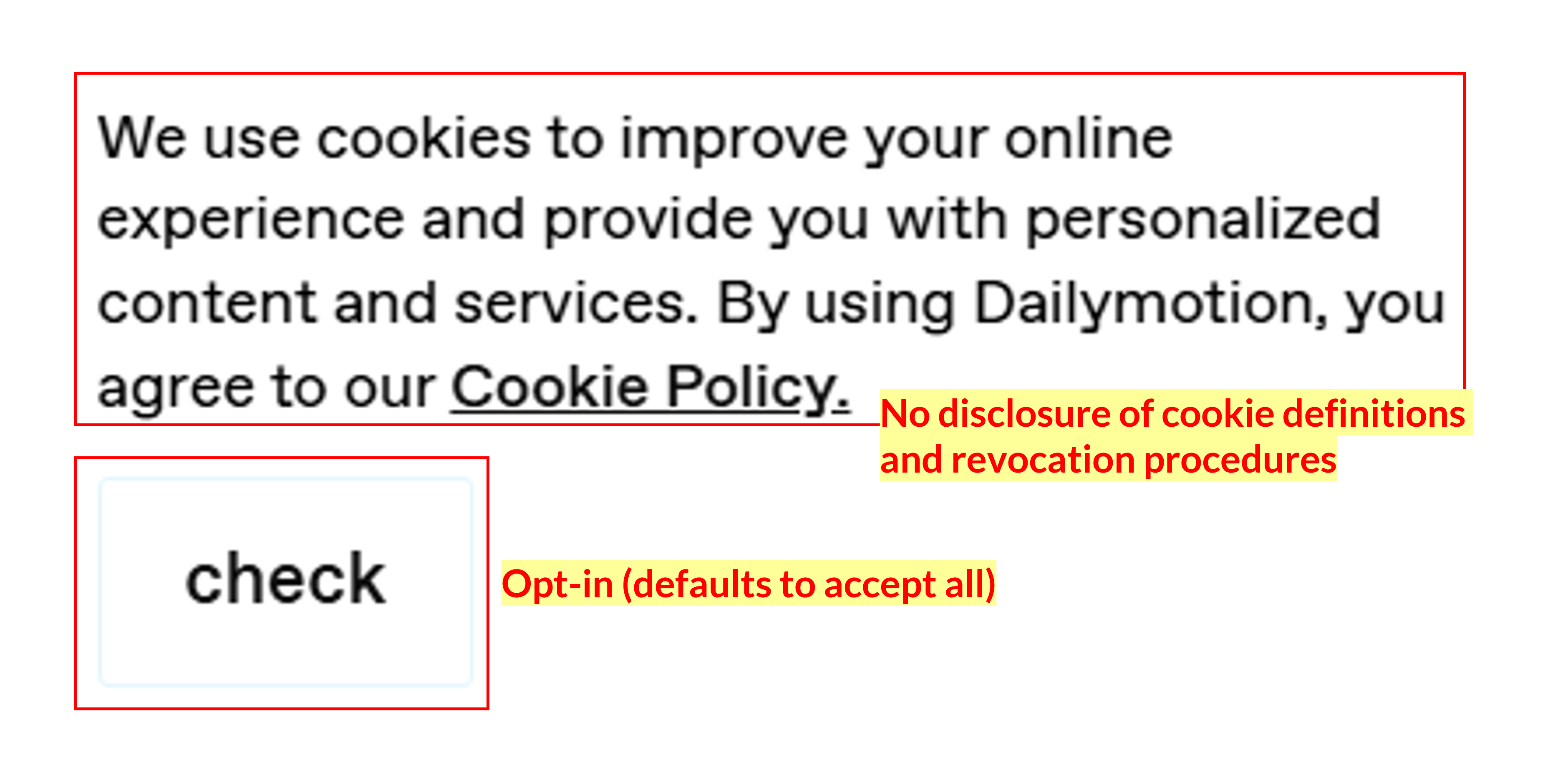}
    \caption{The evolved cookie banner currently on the source site, which includes dark patterns like ambiguous close buttons and the multi-click opt-out (DP19).}
    \Description{The evolved cookie banner currently on the source site, which includes dark patterns like ambiguous close buttons and the multi-click opt-out (DP19).}
    \label{fig:Evolved_banner4_new}
\end{figure}

In this section, we present both the previous and evolved cookie banners from the same source site, as shown in Figures~\ref{fig:Evolved_banner1_old}, \ref{fig:Evolved_banner2_old}, \ref{fig:Evolved_banner3_old}, and \ref{fig:Evolved_banner4_old}. These legacy banners, which are discussed in the article~\cite{kirkman2023darkdialogs}, represent older designs with limited user control and transparency. After conducting a manual examination of the site, we identified the evolved cookie banners, which have significantly changed. The newer designs no longer display the earlier dark patterns, but instead introduce more subtle and legally ambiguous dark patterns, such as ambiguous close options and DP19. These evolved banners, which reflect more recent trends in cookie consent designs, are depicted in Figures~\ref{fig:Evolved_banner1_new}, \ref{fig:Evolved_banner2_new}, \ref{fig:Evolved_banner3_new}, and \ref{fig:Evolved_banner4_new}. This comparison highlights the ongoing evolution of cookie banners and the introduction of new manipulative tactics that continue to affect user consent.

\twocolumn
\section{Summary of Dark Patterns, Description and Respective Legal Violations} 
In Table~\ref{tab:Legal_Mapping}, we present the mappings of evolved dark patterns, including their descriptions, direct regulatory references, and associated legal violations. These mappings offer a comprehensive overview of how modern cookie consent designs exploit user behavior through manipulative tactics. Each dark pattern is carefully analyzed, with a focus on how it violates key principles of user consent and data protection, as outlined in relevant privacy regulations such as the General Data Protection Regulation (GDPR) and the California Consumer Privacy Act (CCPA).
\begin{table}
\centering
\caption{Summary of Dark Pattern Types, Descriptions, and Legal Violations}
\small
\begin{tabular}{>{\bfseries}p{2cm} p{2.5cm} p{3.0cm}}
\toprule
\rowcolor{gray!30}\textbf{Pattern} & \textbf{Description} & \textbf{Regulation} \\
\midrule
DP1 \hspace{3cm} OnlyOptIn & Only Opt-in option on initial Cookie Dialog & GDPR Art. 7(3); EDPB 05/2020: Refusal must be as easy as acceptance \\
\rowcolor{gray!10}
DP2 \hspace{2cm} HighlightedOptIn & Opt-in button highlighted more than Opt-out & GDPR Recital 42, Art. 7(2); EDPB/CNIL: Avoid deceptive design \\
DP3  \hspace{3cm}ObstructWindow & Dialog obstructs the window & GDPR Recital 32, Art. 7(2); EDPB 05/2020: Cookie walls are noncompliant \\
\rowcolor{gray!10}
DP4 \hspace{3cm} ComplexText & Large amount of text on cookie dialog & GDPR Art. 12(1): Information must be clear and accessible \\
DP5 \hspace{3cm} MoreOptions & Multiple layers in cookie dialog & GDPR Art. 7(3); EDPB: Consent withdrawal should not be buried \\
\rowcolor{gray!10}
DP6 \hspace{3cm} AmbiguousClose & Ambiguous Close button in addition to Accept & EDPB 05/2020: Close button not equal to valid consent; CNIL: Intent must match interaction \\
DP7\hspace{3cm}  MultipleDialogs & Multiple distinct cookie dialogs on a page & GDPR Art. 5(1)(a): Multiple dialogs reduce clarity \\
\rowcolor{gray!10}
DP8 \hspace{3cm} PreferenceSlider & At least one slider enabled by default & GDPR Art. 7, Art. 25: Pre-selected options violate freely given consent \\
DP9 \hspace{3cm} CloseMoreCookies & Clicking Close leads to more cookies set & GDPR Art. 6(1)(a); EDPB 05/2020: Passive acceptance is invalid \\
\rowcolor{gray!10}
DP10\hspace{3cm}  OptOutMoreCookies & More cookies set despite clicking Opt-out & GDPR Art. 5(1)(a–f): Tracking after opt-out violates purpose limitation \\
\rowcolor{gray!10}
DP12 \hspace{3cm} PurposeInfoDisplay & No cookie purpose info on the first page & GDPR Art. 13(1)(c): Purpose must be stated \\
DP13\hspace{3cm}  OptOutPricing & Pricing based on opt-out decision & GDPR Art. 5(1)(b); CCPA §1798.125: Price discrimination must be justified \\
\rowcolor{gray!10}
DP14 \hspace{3cm} ConsentRevocationPossible & Consent revocation not possible & GDPR Art. 7(3); Recital 42: Consent must be withdrawable \\
DP15 \hspace{3cm} RevocationHard & Consent revocation is hard & GDPR Art. 7(3), Recital 42: Revocation must be easy and accessible \\
\rowcolor{gray!10}
DP16 \hspace{3cm} PreConsentCookies & Pre-consent cookie loading & ePrivacy Art. 5(3); CJEU Planet49 (C-673/17); GDPR Recital 32: Consent required before cookies \\
DP17\hspace{3cm}  LegalAmbiguity & Unclear legal basis for cookies & GDPR Art. 6(1), Art. 13(1)(c), Recital 39: Clear lawful basis must be stated \\
\rowcolor{gray!10}
DP18\hspace{3cm}  FakeOptOut & Default rejection is non-functional & GDPR Art. 7(4), Recitals 42–43; EDPB 05/2020: Refusal must be genuine \\
DP19 \hspace{3cm} MultiClickOptOut & Takes more clicks to Opt-in or Opt-out & EDPB Cookie Banner Taskforce (2023); CNIL 2021: Reject must be as easy and visible as Accept \\
\bottomrule
\end{tabular}
\label{tab:Legal_Mapping}
\end{table}

\end{document}